\documentclass[12pt,preprint]{aastex}
\usepackage{psfig}

\newcommand{\kms}{km~s$^{-1}$}
\newcommand{\sunn}{$_{\odot}$}
\newcounter{qub}
\begin{document}

\title{
Strong Emission Line H{\sc ii} Galaxies in the 
Sloan Digital Sky Survey. \\
I. Catalog of DR1 Objects with Oxygen Abundances \\
from $T_{\rm e}$ Measurements
}

\author{Alexei Y. Kniazev\altaffilmark{1,2,3},
Simon A. Pustilnik\altaffilmark{2,3},
Eva K. Grebel\altaffilmark{1,4},
Henry Lee\altaffilmark{1,5},
and Alexander G. Pramskij\altaffilmark{2,3}}

\email{kniazev@mpia.de, sap@sao.ru, grebel@astro.unibas.ch,
       hlee@astro.umn.edu, pramsky@sao.ru}

\altaffiltext{1}{Max-Planck-Institut f\"{u}r Astronomie, K\"{o}nigstuhl 17, D-69117 Heidelberg, Germany}
\altaffiltext{2}{Special Astrophysical Observatory, Nizhnij Arkhyz,
		 Karachai-Circassia, 369167, Russia}
\altaffiltext{3}{Isaac Newton Institute of Chile, SAO Branch}
\altaffiltext{4}{Present address: Astronomisches Institut der
  Universit\"at Basel, Venusstrasse 7, CH-4102 Binningen, Switzerland.}
\altaffiltext{5}{Present address: Dept. of Astronomy, Univ. of
  Minnesota, 116 Church St. S.E., Minneapolis, MN, 55455 USA.}

\begin{abstract}
We present the first edition of the \underline{\bf S}DSS
\underline{\bf H}{\sc ii}-galaxies 
with \underline{\bf O}xygen abundances \underline{\bf C}atalog (SHOC),
which is a listing of strong emission-line galaxies (ELGs) from
the Sloan Digital Sky Survey (SDSS).
Oxygen abundances have been obtained with the classic 
$T_{\rm e}$-method. We describe the method exploiting the
SDSS database to construct this sample.
The selection procedures are described and discussed in
detail, as well as some problems encountered in the process of
deriving reliable emission line parameters.
The method was applied to the SDSS Data Release 1 (DR1).
We present 612 SDSS
emission-line galaxies (624 separate SDSS targets in total),
for which the oxygen abundances 12+$\log$(O/H) have
r.m.s. uncertainties $\le\,$0.20 dex. 
The subsample of
263 ELGs (272 separate SDSS targets) have an
uncertainty $\le\,$0.10 dex, while 459 ELGs (470 separate SDSS targets)
have an uncertainty $\le$0.15 dex.
The catalog includes the main parameters of all selected ELGs, the
intensities and equivalent widths of hydrogen and oxygen
emission lines, as well as oxygen abundances with their uncertainties.
The information on the presence of Wolf-Rayet blue and/or red
bumps in 109 galaxies is also included.
With the use of combined $g,r,i$ SDSS images we performed visual
morphological classification of all SHOC galaxies.
461 galaxies ($\sim$75\%) are classified as confident or probable blue
compact galaxies (BCG/BCG?),
78 as irregular ones,
20 as low surface brightness galaxies (LSBG),
10 as obviously interacting
and 43 as spiral galaxies.
In creating the catalog, 30 narrow line AGN and 69 LINERs were
also identified; these are also presented apart of the main catalog.
We outline briefly the content of the catalog, and the prospects of its
use for statistical studies of the star formation and chemical evolution
issues. 
Some of these studies will be presented in the forthcoming paper.
Finally, we show that the method presented by \citet{Kniazev03a}
for calculating O$^+$/H$^+$ using intensities of the
[O\,{\sc ii}] $\lambda$7320,7330 \AA\ lines for SDSS emission-line
spectra in the absence of [O\,{\sc ii}] $\lambda$3727 \AA\ line
appears to yield reliable results over a wide
range of studied oxygen abundances: $7.10 < 12+\log$(O/H) $ < 8.5$.
\end{abstract}


\section{Introduction}

The heavy element abundances of gas-rich galaxies and their
gas-mass fractions are the main parameters characterizing their global
evolution \citep[ e.g.,][]{Pagel97,Matteucci01}.
These can be related to both the global properties of galaxies
\citep[e.g.,][]{Dalcanton97,GGH03}
and the membership of galaxies in various elements of large-scale
structure \citep[e.g.,][]{Popescu96,Grogin00,Pustilnik02b,Vilchez03,Lee03}.
The knowledge of the metallicity for large homogeneously selected galaxy
samples would allow us to address various issues of galaxy chemical evolution
on a good statistical basis.
In particular, the possible difference in the chemical evolution rate in
the various types of galaxy environments can be systematically examined.
Having metallicities for galaxy samples at redshifts of, e.g.,
$z \sim 0.3$ and in the nearby Universe, one can directly probe the
chemical evolution of gas-rich galaxies over timescales of several Gyr.
In addition, estimates of stellar mass from galaxy photometry and
of neutral gas mass from H{\sc i} measurements bring new opportunities
to confront the predictions of modern chemical evolution models
with the observed properties of a large sample of galaxies.

To address the question of metallicity distributions in gas-rich galaxies and
to understand possible relations between metallicity and other galaxy
parameters, one can use several large ELG samples such as the results of
University of Michigan (UM), Tololo and Calan-Tololo
\citep{Smith76,McAlpine77,Salzer89},
Case \citep{Pesch82,Salzer95,Ugryumov98},
Second Byurakan Survey  \citep[SBS;][]{Markarian83,Izotov93},
Heidelberg Void \citep{Popescu96},
KPNO International Spectral Survey \citep[KISS;][]{Salzer00,Melbourne02},
Hamburg/SAO Survey for Emission-Line Galaxies
\citep[HSS-ELG;][]{Ugryumov99,Pustilnik00}
and Hamburg/SAO Survey for Low Metallicity Galaxies
\citep[HSS-LM;][]{Ugryumov03}.

Of special interest for such samples are H{\sc ii} galaxies
and their most
prominent representatives -- Blue Compact Galaxies (BCGs).
BCGs are gas-rich objects with typical total
masses lower than 10$^{10}~M$\sunn, have low metallicities
in the range 1/15$ \leq$ Z/Z\sunn\ $\leq$ 1/3, and form stars
at noticeably non-stationary rates.
Previously, samples of H{\sc ii} galaxies with reliably known
metallicities (i.e., derived with the $T_{\rm e}$ method)
were obtained through high S/N spectroscopy of
strong-line ELGs, selected from the surveys cited above.
However, well-selected gas-rich galaxy samples with reliable
metallicity determinations are still quite small.
Currently, it is possible to estimate that we have no more
than $\sim$200 galaxies selected from the different samples
with measured classic $T_{\rm e}$ method
oxygen abundances with an accuracy better than 0.1 dex.
This is related to the weakness of the key temperature-sensitive
line [O\,{\sc iii}]$\lambda$4363, used in the classic $T_{\rm e}$ method to
derive oxygen abundances with r.m.s. uncertainties of 0.01--0.1 dex.
As well, many galaxies from these samples often have poor
photometry, and, apart from the KISS survey, selection criteria are
not well-defined in terms of apparent magnitude.
Nevertheless, the accumulated data on low-mass galaxies give
important clues about the metallicity distribution and indicate
correlations with other galaxy global parameters.

In particular, these surveys have uncovered a significant number of
extremely metal-poor galaxies (XMPGs)\footnote{Another name for these
galaxies is extremely metal-deficient galaxies (XMDs).} 
with $Z \leq 1/20~Z$\sunn. 
Some XMPGs are similar to the well-known I~Zw~18 \citep{SS72} 
and SBS 0335--052 \citep{Izotov90},
which are candidates for young galaxies in the nearby Universe.
These are probably the best local analogs of young low-mass
galaxies which formed at high redshifts.
Despite the paucity of such galaxies
their systematic study can advance significantly the understanding of
the details of galaxy formation and their early evolution.
Therefore, it is important to have an effective means of enlarging
substantially the number of XMPGs.

Besides the great interest in understanding the details of star
formation, massive-star (including Wolf-Rayet stars) evolution and their
interaction with the interstellar medium at very low metallicities,
there are several other important directions related to the studies of
H{\sc ii}/BCG metallicities in general.
For example, with a larger ELG
sample with abundances measured by the classic $T_{\rm e}$ method we
can improve the calibration of the empirical methods
\citep[e.g.,][]{Pagel,McGaugh91,Pilyugin01,Pilyugin03,Denicolo},
which provide a broad picture about
the range of oxygen abundances for ELGs in general.

Therefore, it is natural to look for new opportunities offered by
the Sloan Digital Sky Survey \citep[SDSS;][]{York2000}.
Owing to its homogeneity, area coverage, spectral resolution,
and depth, the SDSS provides an
excellent means of creating a large flux-limited sample of H{\sc ii} galaxies
with heavy element abundances derived with the classic $T_{\rm e}$
method.

The SDSS consists of an
imaging survey in five photometric bands \citep{SDSS_phot,Gunn98,Hogg01},
as well as a follow-up spectroscopic survey of
a magnitude-limited sample of galaxies
\citep[mainly field galaxies brighter than $r = 17\fm77$;][]{Strauss02}
and a color-selected sample of quasars \citep{QSO02}.
An automated image-processing system detects
astronomical sources and measures their photometric and astrometric
properties \citep{Lupton01,EDR02,SDSS_phot1,Pier03}
and identifies candidate galaxies and quasars for
multi-fibre spectroscopy.  The samples of galaxy and quasar
candidates include a substantial number of emission-line galaxies.
The spectra are automatically reduced and wavelength- and flux-calibrated
\citep{EDR02,DR1}.

The SDSS spectral data have already been used
in a number of galaxy studies
\citep[e.g.,][]{Bernardi03,Eisen02,Goto03,Kauffmann03,Kniazev03a,Stas03}.
The paper presented here will extend the possibilities of using
SDSS ELG spectra for the statistical studies of galaxy metallicities.

In this paper we describe the method used to extract from the SDSS
database strong-line ELGs, which are suitable for determining
oxygen abundances with the classic $T_{\rm e}$ method, i.e.,
the temperature-sensitive [O$\;${\sc iii}] $\lambda$4363~\AA\ line.
The original galaxy sample is obtained from the SDSS DR1,
which is briefly described in Section~\ref{txt:Selection}.
The application of the developed pipeline yields the list of ELGs with
H{\sc ii}-type spectra.
In the same section the procedure of the ELG selection is described in detail.
The method used to estimate the physical conditions in 
H{\sc ii} regions of the galaxies studied and their element
abundances is described in Section~\ref{Method}.
In the same section we outline a number of problems encountered
while using the SDSS spectral data and the ways to
resolve them. 
In Section~\ref{Tests_OH} we check the quality of the oxygen
abundance determination.
The catalog of all selected ELGs along with their main parameters, emission
line data, and the derived oxygen abundances is presented in
Section~\ref{Results}.
The results are presented in Section~\ref{Discussion}.
We conclude with the key results of this paper in Section~\ref{Summary}.
We adopt here the Hubble constant $H_\mathrm{0}$ = 75~\kms~Mpc$^{-1}$.

\section{Sample selection}
\label{txt:Selection}

\subsection{DR1 spectral data}

The SDSS uses two fiber-fed double-spectrographs to
measure spectra of objects in the range
from 3800 \AA\ to 9200 \AA\ with 
spectral resolution $R \sim 1800$.
Each spectrograph handles
320 fibers (which we call a half-plate), and the two halves are
reduced independently.
For a single pointing, each observed plate contains 640 fibers,
yielding 608 spectra of galaxies, quasars, and stars, and 32 sky
spectra \citep{Bla03}. The fibers have a diameter of $3''$.
The instrumental resolution and pixel scale are close to
constant in logarithmic wavelength rather than linear wavelength.
The flux calibration procedure is summarized in
\citet[Sections~3.3 \& 4.10.1]{EDR02}
and will be fully described by Schlegel et al. (in preparation).
The flux calibration is imposed on each plate by a set of 
eight spectrophotometric standard stars, chosen by color to be
F-subdwarf stars. 

For our work we used the reduced spectra from the DR1 \citep{DR1}
database which cover a total area on the sky of $\sim$1360 deg$^2$.
All these spectra have been copied from
the official DR1 web-site (see {\tt http://www.sdss.org/dr1/} for details)
as two-dimensional FITS-files; one file for each spectral plate contains
one spectrum per row.
Also, the FITS file containing photometric and some spectral information
for 133,996 DR1 galaxies with observed spectra was copied and used as
a primary database in our work. 

\subsection{Measurement of lines}
\label{txt:lines}

For the analysis of the SDSS spectra we used our own software for
emission-line data,
created for the HSS-ELG and for HSS-LM projects \citep{Ugryumov01,Ugryumov03}.
This software is based on the MIDAS\footnote{MIDAS is an acronym
for the Munich Image Data Analysis System, which is a package
developed for the European Southern Observatory.} Command Language
programs and was adapted especially for the requirements of
the SDSS spectral data. The programs dealing with the fitting
of emission/absorption line parameters are based on the MIDAS
programs SET/FIT, FIT/IMAGE, COMPUTE/FIT and SAVE/FIT
from the {\tt FIT} package \citep{MidasA}.
The line fitting was based on the Corrected Gauss-Newton method.
Every line in the reference list was fitted as a single Gaussian 
superimposed on the continuum-subtracted spectrum.
Some lines were fitted simultaneously as
a blend of two or more Gaussians features. 
For the current work, Gaussian blends were obtained for
the H$\alpha$ 6563 \AA\ and 
[N~{\sc ii}] $\lambda\lambda$6548,6584 \AA\ lines,
[S~{\sc ii}] $\lambda\lambda$6716,6731 \AA\ lines, and
[O~{\sc ii}] $\lambda\lambda$7320,7330 \AA\ lines.

The continuum was determined with the help
of the algorithm described in detail by \citet{Sh_Kn_Li_96}.
This algorithm initially had been developed in the Special Astrophysical
Observatory of the Russian Academy of Sciences
for the reduction package of one-dimensional radio data aimed
at the detection of weak sources.
It was also successfully used in the reduction of KISS data \citep{Akn97},
where it showed very robust results even for the objective prism spectra
with the total length of only forty points.
The clipping procedure of the algorithm deals with the noise level
of a spectrum that in the general case changes from point to point;
the noise level is thus defined as a function $\sigma(\lambda)$.
By the definition of the algorithm, the fitted continuum also has
an uncertainty $\sigma(\lambda)$ which is added to the uncertainty of the
measured line intensity of the studied spectrum. 
Our continuum fit algorithm does not produce a continuum noise estimate.
For the latter, we used the Absolute Median Deviation (AMD) estimator,
$med (|x-\overline{x}|)$, where $\overline{x}$ is the mean of the input
distribution \citep{Korn68}.
If $\sigma_{\rm n}$ is the standard deviation for a normal
distribution, then the standard deviation of the AMD estimator is
$\sigma_{\rm AMD} \approx 0.674\:\sigma_{\rm n}$.
Thus, the final noise estimate, $\sigma$, should be corrected 
as $\sigma = \sigma_{\rm AMD} / 0.674$.
The AMD algorithm is a fairly fast robust procedure and has
been used for the stream data reduction system
of observations with the RATAN--600 radiotelescope \citep{Erukhimov88}.
Analysis of the algorithm, performed for  Gaussian noise plus
noise spikes with Poisson distribution,
showed good stability of this estimator: the estimated value 
does not depend on the intensity of noise spikes.
This was a reason to use this estimator for spectra with strong
emission lines.

The quoted errors in the line intensities $\sigma_{\rm tot}$ include
two components.
First, $\sigma_{\rm f}$ is the fitting error from the MIDAS program
{\tt FIT/IMAGE} and is related to Poisson statistics of line photon flux.
Second, $\sigma_{\rm c}$ is the error resulting from the creation of
the underlying continuum (calculated using the AMD estimator), which
is the largest error contributor for faint lines.
So, the final error is calculated as:
\begin{equation}
\sigma_{\rm tot} = \sqrt{\sigma_{\rm f}^2 + \sigma_{\rm c}^2}
\label{eq:lines}
\end{equation}

\subsection{Selection steps}
\label{txt:sel_steps}

To  simplify the work with the  entire sample of $\sim$134,000
spectra, we divided our selection procedure into two steps:

1. Fast measurements of the equivalent widths (EWs) of the
strongest emission Balmer lines H$\alpha$, H$\beta$ and H$\gamma$
for all objects allowed us to preselect a subsample
with the strongest emission lines.
With this step, we chose the Balmer lines
instead of the [O~{\sc iii}] $\lambda$4959,5007~\AA\  lines.
The reason is related to the  steps described below, in which we
plan to use the complete subsample of H{\sc ii} galaxies limited by
the value of EW(H$\beta$), since the latter can serve as an indicator
of the age of the starburst \citep[e.g.,][]{SV98}.

\begin{figure}
    \begin{center}
    \epsscale{1.0}
\plotone{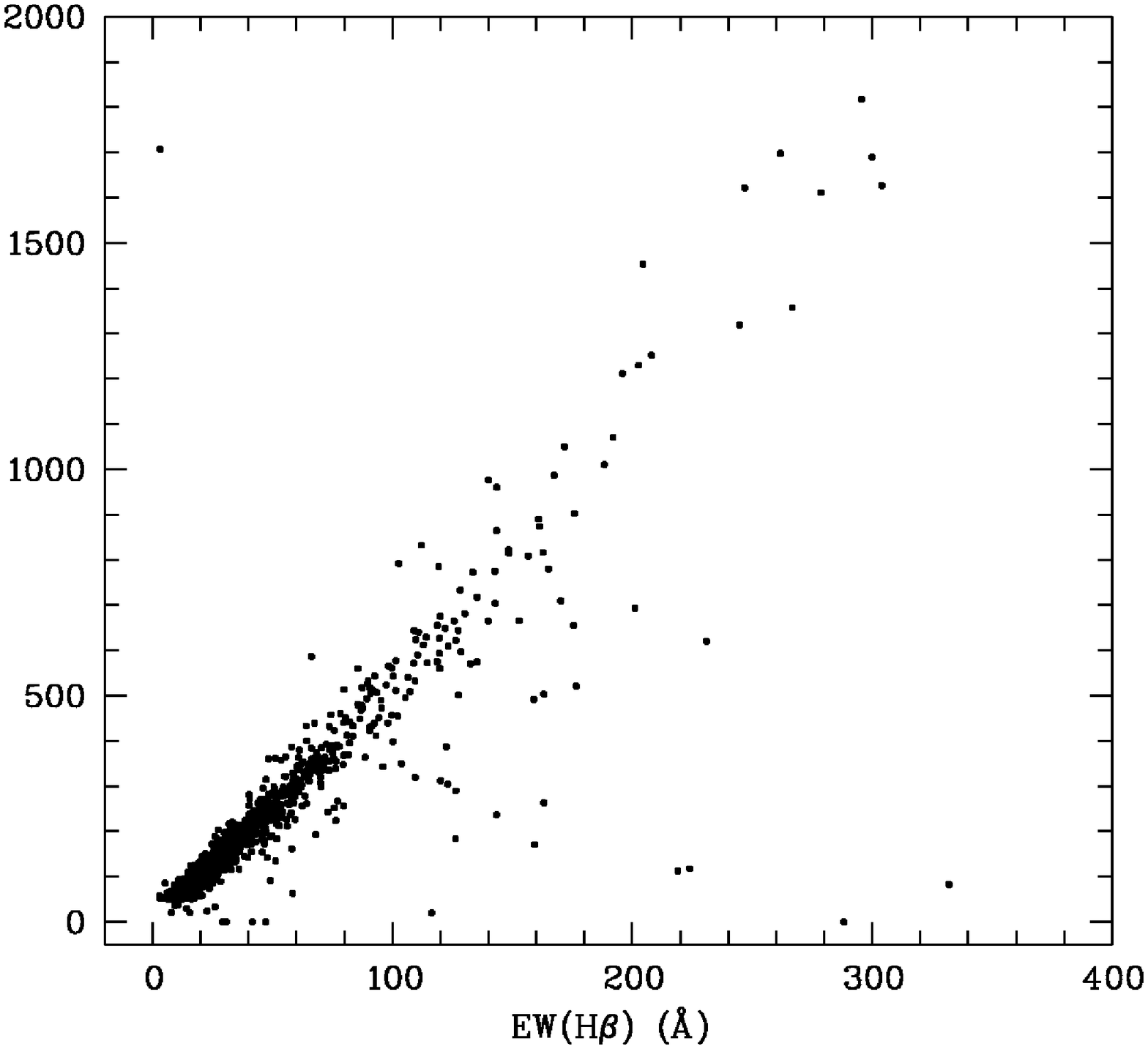}
    \caption{The relation between the EWs of the H$\beta$ and H$\alpha$ lines
       for all preselected ELGs ($\sim$5000 spectra).
       The truncation of the H$\alpha$ line is seen in all spectra
	whose points are located significantly below the main locus.
    \label{fig:truncation}}
    \end{center}
\end{figure}

We found that either the SDSS pipeline for the DR1 data
truncated some fraction of the strongest emission lines in the galaxy
spectra, or that the lines themselves were saturated.
In Fig.~\ref{fig:truncation} we show, as an example, the plot
of EW(H$\alpha$) versus EW(H$\beta$) for our sample.
For those data points that are
located in this Figure below the "main sequence",
the H$\alpha$ line was truncated during the pipeline
reduction or/and was saturated in the observations.
We found that this problem can be overcome at the primary selection
step, if we use simultaneously the three strongest Balmer lines
(H$\alpha$, H$\beta$ and H$\gamma$) to preselect objects with the
strongest oxygen emission. 
That is, the galaxy was considered to pass the preliminary selection
procedure if one of the following EW thresholds was observed:
\begin{equation}
{\rm EW}(H\gamma) \ge 6\;{\rm \AA} \,||\,
{\rm EW}(H\beta) \ge 20\;{\rm \AA} \,||\,
{\rm EW}(H\alpha) \ge 50\;{\rm \AA}
\end{equation}

Altogether, $\sim$5000 spectra were preselected with these criteria.
The principal parameter of the imposed selection criteria is based
on the threshold value of EW(H$\beta$). The two other thresholds are
related to the former approximately through the theoretical Balmer decrement.
The threshold value of the EW(H$\beta$) selection criterion
was motivated by the requirement to have the  measurable
weak [O\,{\sc iii}] $\lambda$4363 \AA\ emission line, necessary
for a direct calculation of the electron temperature, $T_e$.
This line normally is fainter than the [O\,{\sc iii}] $\lambda$5007 \AA\ line
by a factor of 200 to 40 in the $T_{\rm e}$ range 10,000~K to 20,000~K
\citep{Aller84}.
The intensities of [O\,{\sc iii}] $\lambda$5007 \AA\ and H$\beta$ in H{\sc ii}
galaxies statistically have
an average ratio [O\,{\sc iii}]/H$\beta$ of approximately
$4 \pm 1$ \citep[e.g.,][]{Ugryumov03}.
This translates to a  H$\beta$/[O\,{\sc iii}] $\lambda$4363 \AA\
intensity ratio of 50 to 10.
Since the faintest measurable [O{\sc iii}] $\lambda$4363 \AA\
lines in the spectra
of the SDSS ELGs are found to have the EW $\sim$0.4 \AA, this implies that
the primary selection criterion for the related H$\beta$-line should
be EW(H$\beta$)$\gtrsim$20~\AA.
The standard Balmer ratio I($H\alpha$)/I($H\beta$) is 2.88 at 10,000~K.
The observed value will be larger, if any extinction is present.
The standard Balmer ratio I($H\gamma$)/I($H\beta$) is 0.47 at 10,000~K.
Thus, our imposed selection criteria for the EW(H$\alpha$) and
EW($H\gamma$) are somewhat softer relative to that for EW(H$\beta$).

With the models of \citet{Veilleux87} and \citet{Baldwin81}
we selected 99 narrow-line active galactic nuclei (AGN)
and low-ionization nuclear emission-line region galaxies 
(LINERs; \citealp{Heckman80}).
AGNs and LINERs are ionized by a non-thermal power law continuum
and/or shock heating; thus, these galaxies were removed from
the list.
A diagram used for the classification and selection
is shown in Fig.~\ref{fig:class_diag}.

\begin{figure}
    \begin{center}
    \epsscale{1.0}
\plotone{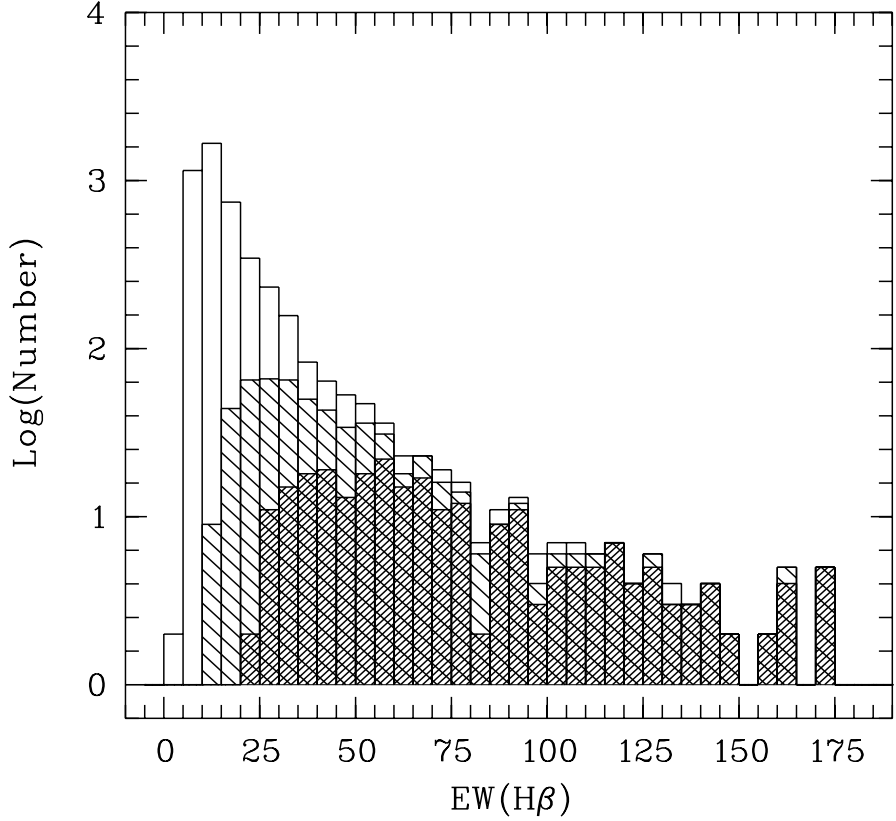}
    \caption{
 Distributions of EW(H$\beta$):
 all preselected ELGs (open bins, 4773 spectra),
 the final selected sample with an accuracy of $\log$(O/H) $\le$ 0.2 dex
 (hashed bins, 624 spectra) and the sample with an accuracy of $\log$(O/H)
 $\le$ 0.1 dex (double-hashed bins, 272 spectra).
 17 galaxies are outside the plot region, having EW(H$\beta$) up to 356~\AA.
 All of them have an accuracy of $\log$(O/H) $\le$ 0.1 dex.
 All AGNs and repeated observations are not shown in the plot.
    \label{fig:Hb_distrib}}
    \end{center}
\end{figure}

2. 
We measured all emission line intensities and calculated 
the chemical abundances for all objects in the preselected subsample.
The method used for the calculations is described in
Section~\ref{method_abun}.
Altogether, 638 spectra were finally selected by imposing the
criterion of the accuracy of the oxygen abundance $\le$~0.2 dex.
We found that among these 638 spectra, 28 belonged to 14 SDSS targets
which were observed twice.
EW(H$\beta$) distributions for the preselected and final samples are
shown in Fig.~\ref{fig:Hb_distrib}.
The main parameters of the galaxies from the final selected sample
are summarized in Table~\ref{tab:General}
and measured lines for selected galaxies are shown in 
Tables~\ref{t:Intens} and \ref{t:Intens_corr}.
A complete description of these Tables is presented in
Section~\ref{Results}.

\section{Physical conditions and heavy element abundances}
\label{Method}

\subsection{The Method }
\label{method_abun}

The measured emission line intensities $F(\lambda)$ were corrected both for
reddening and  for the effects of underlying stellar absorption
following the procedure described in detail by \citet{Izotov94}.
We have adopted  an iterative procedure
to derive simultaneously both the extinction coefficient
C(H$\beta$) and the absorption equivalent width for the hydrogen lines from
the equation \citep{Izotov94}:
\begin{eqnarray}
\frac{I(\lambda)}{I({\rm H}\beta)}& = \frac{{\rm EW}_e(\lambda)+{\rm EW}_a(\lambda)}{{\rm EW}_e(\lambda)} 
\frac{{\rm EW}_e({\rm H}\beta)}{{\rm EW}_e({\rm H}\beta)+{\rm EW}_a({\rm H}\beta)} \cdot \nonumber \\
			    & \frac{F(\lambda)}{F({\rm H}\beta)} \exp{[C({\rm H}\beta)f(\lambda)]},
\label{eq:CHb}
\end{eqnarray}
where I($\lambda$) is the intrinsic line flux and F($\lambda$) is the
observed line flux corrected for atmospheric extinction.
EW$_e$($\lambda$) and EW$_a$($\lambda$) are the equivalent widths of
the observed emission line and of the underlying absorption line,
respectively. 
$f(\lambda)$ is the reddening function, normalized at
H$\beta$. The theoretical ratios from \citet{Brocklehurst71} for
the intrinsic hydrogen line intensity ratios for estimated
electron temperature were used.
For lines other than  hydrogen EW$_a$($\lambda$)=0 and
equation (\ref{eq:CHb}) reduces to 
\begin{equation}
\frac{I(\lambda)}{I({\rm H}\beta)} = \frac{F(\lambda)}{F({\rm H}\beta)} \exp{[C({\rm H}\beta)f(\lambda)]}.
\label{eq:CHb1}
\end{equation}

To derive element abundances of oxygen, we  use the
standard two-zone model and  the
method from \citet{Aller84}, and also follow the procedure
described by \citet{Izotov94,TIL95,Izotov97}, and \citet{IT99}.
The electron temperature T$_{\rm e}$ is known to be different in high- and
low-ionization H{\sc ii} regions \citep{Stas90}.
The procedure determines T$_{\rm e}$(O {\sc iii}) from the
[O {\sc iii}]$\lambda$4363/($\lambda$4959+$\lambda$5007)
ratio using the five-level atom model \citep{Aller84}
and the electron density N$_{\rm e}$(S {\sc ii}) from the
[S {\sc ii}]$\lambda$6717/$\lambda$6731 ratio.
The minimum value of N$_{\rm e}$(S {\sc ii})
was set to be 10 cm$^{-3}$.
To derive the O$^{+}$ electron temperature, the relation between
T$_{\rm e}$(O {\sc ii}) and T$_{\rm e}$(O {\sc iii}) from a fit
by \citet{Izotov94} to photoionized H{\sc ii} models by \citet{Stas90}
was used:
\begin{equation}
t_e({\rm O\;II})=0.243+
t_e({\rm O\;III}) \left[ 1.031 - 0.184\,t_e({\rm O\;III}) \right],
\label{eq:tOII}
\end{equation}
where t$_{\rm e}$ = T$_{\rm e}$/10$^4$.

For oxygen, the following expression for the total abundance was used:
\begin{equation}
\frac{{\rm O}}{{\rm H}} = \frac{{\rm O}^{+}+{\rm O}^{++}}{{\rm H}^+}.
\label{eq:ICFO}
\end{equation}

All uncertainties in the measurements of line intensities,
continuum level, extinction coefficients, and Balmer absorption
equivalent widths
have been propagated in the calculations of oxygen abundance, and are
accounted for in the accuracies of the presented element abundances.
This is detailed in the algorithm steps shown below:\\
(1). All measured line intensities with the uncertainties specified with
equation~(\ref{eq:lines}) were read and recalculated relative to the
intensity of the H$\beta$ line
${\rm I(\lambda)_{\rm 1}} = {\rm I(\lambda)}/{\rm I(H\beta)}$.
New errors were calculated as
\begin{equation}
\label{equ:new}
\sigma({\rm \lambda})_{\rm 1} = {\rm I(\lambda)_{\rm 1}} \cdot \sqrt{\Big(\frac{\rm \sigma(\lambda)}{\rm I(\lambda)}\Big)^2 + \Big(\frac{\rm \sigma(H\beta)}{\rm I(H\beta)}\Big)^2},
\end{equation}
(2). Using the system of equations~(\ref{eq:CHb}) for
hydrogen lines,
the extinction coefficient C(H$\beta$)$\pm \delta$C(H$\beta$) and
the equivalent width of underlying absorption in Balmer hydrogen lines
EW(abs)$\pm \delta$EW(abs) were calculated.\\
(3). New relative intensities ${\rm I(\lambda)_{\rm 2}}$ for 
hydrogen lines were calculated using EW(abs) from step (2) so that:
${\rm I(\lambda)_2} = {\rm I(\lambda)_1} \cdot ({\rm EW(\lambda) +
EW(abs)})/{\rm EW(\lambda)}$.
The errors $\sigma({\lambda})_{\rm 2}$ for these lines
were recalculated using the determined $\delta$EW(abs).\\
(4). New relative intensities ${\rm I(\lambda)_{\rm 3}}$ for all lines 
were calculated with equation~(\ref{eq:CHb1}) using the C(H$\beta$)
value determined and the reddening function $f(\lambda)$ from
\citet{Whitf58}.  
\cite{Izotov94} gave an approximate reddening function in the entire
spectral region as 
\begin{equation}
f(\lambda)= 3.15854\cdot10^{-1.02109\lambda}-1,
\label{eq:flambda}
\end{equation}
where $\lambda$ is expressed in units of $\mu$m.
Final errors $\sigma({\lambda})_{\rm 3}$ were recalculated using 
determined $\delta$C(H$\beta$) values\footnote{%
Because the theoretical ratios for the hydrogen lines used in
equations~(\ref{eq:CHb}) depend on the electron temperature T$_{\rm e}$,
which can be determined only after step 4 has finished, steps 2
through 4 were performed iteratively until the results converged,
after which the final errors were calculated. 
}.
All these relative intensities with their errors
are presented in Table~\ref{t:Intens_corr} and are used for
calculation of temperatures, densities and oxygen abundances.
Calculated oxygen abundances for selected galaxies are presented
in Table~\ref{t:Abun}.

\subsection{[O {\sc ii}]~$\lambda$3727~\AA\ detection problem}
\label{line3727}

Since the SDSS spectra are acquired in the range $3800-9000$~\AA, the
line [O~{\sc ii}]~$\lambda$3727~\AA\ is out of the range (or is very
close to the edge) for objects with redshifts $ \lesssim$~0.024$-$0.025.
Therefore, for such galaxies the determination of the
O$^+$/H$^+$ abundance by the standard method, for which the intensity of
[O~{\sc ii}]~$\lambda$3727~\AA\ is used,  
is impossible.
However, for most of the SDSS spectra of H{\sc ii} galaxies this
problem can be overcome by a small modification of the
standard method. 
As shown by \citet[][ p.130]{Aller84}, the value of O$^+$/H$^+$ can be
equally well calculated from the intensities of the auroral lines
[O\,{\sc ii}] $\lambda$7320,7330 \AA.
The necessary auxiliary quantities for the auroral lines (L$_A$), as well
as those for the nebular ones (L$_N$), used in the respective formula
for the calculation of ionic concentrations, are tabulated
in Table 5-5b of \citet{Aller84}.
Both methods should give the same value of O$^+$/H$^+$
\citep[see,][for all details]{Aller84},
but since the
total intensities of the [O\,{\sc ii}] $\lambda$7320,7330 \AA\ lines
are many times lower than that of the [O\,{\sc ii}] $\lambda$3727 \AA\ 
line, application of the auroral line method is restricted to SDSS
spectra with sufficiently high signal-to-noise ratio.
This method was first employed by \citet{Kniazev03a}\footnote{%
In this work [O\,{\sc ii}] $\lambda$7320,7330 \AA\ lines were
used only if the line
[O{\sc ii}] $\lambda$3727 \AA\ was not detected in the spectra.}.
We show in Section~\ref{Tests_OH} that this method gives
reliable results over the entire range of abundances studied here.

\subsection{Strong lines truncation}
\label{Truncation}

One of the problems, already mentioned in Section~\ref{txt:sel_steps},
was a significant truncation  of the line emission for e.g.,
H$\alpha$, H$\beta$, [O~{\sc iii}]$\lambda\lambda$4959,5007 \AA, 
in a fraction of spectra obtained.
Hereafter we assume that a studied line {\it 1} with flux I$_{1}$ is
truncated if the measured flux ratio of lines {\it 1} and {\it 2}
satisfy the condition:
I$_{1}$/I$_{2}$ is lower than the standard value of this ratio by
more than 5$\sigma_{\rm ratio}$. $\sigma_{\rm ratio}$ is calculated as:
\begin{equation}
\label{equ:truncation}
\sigma_{\rm ratio} = \frac{\rm I_{1}}{\rm I_{2}} \cdot \sqrt{\Big(\frac{\sigma_{1}}{\rm I_{1}}\Big)^2 + \Big(\frac{\sigma_{2}}{\rm I_{2}}\Big)^2},
\end{equation}
where $\sigma_{1}$ and $\sigma_{2}$ are observational uncertainties of
I$_{1}$ and I$_{2}$.
The objects most likely affected are those that yield the most
accurate element abundances, because their H{\sc ii} regions often have
the strongest lines.
There are two possible explanations for the truncation:
(1) saturation of the strong emission lines,
and (2) problems with the standard SDSS pipeline.
The latter may result from the procedure used to clean images of
cosmic ray contamination.
The fraction of truncated spectra in the DR1 database is significantly
reduced in comparison to the Early Data Release. For example, the
17 selected spectra from the EDR database with truncated H$\alpha$
line emission all appeared unaffected in the DR1 database, which
supports the second suggestion about pipeline problems.

\begin{figure}
    \begin{center}
    \epsscale{1.0}
\plotone{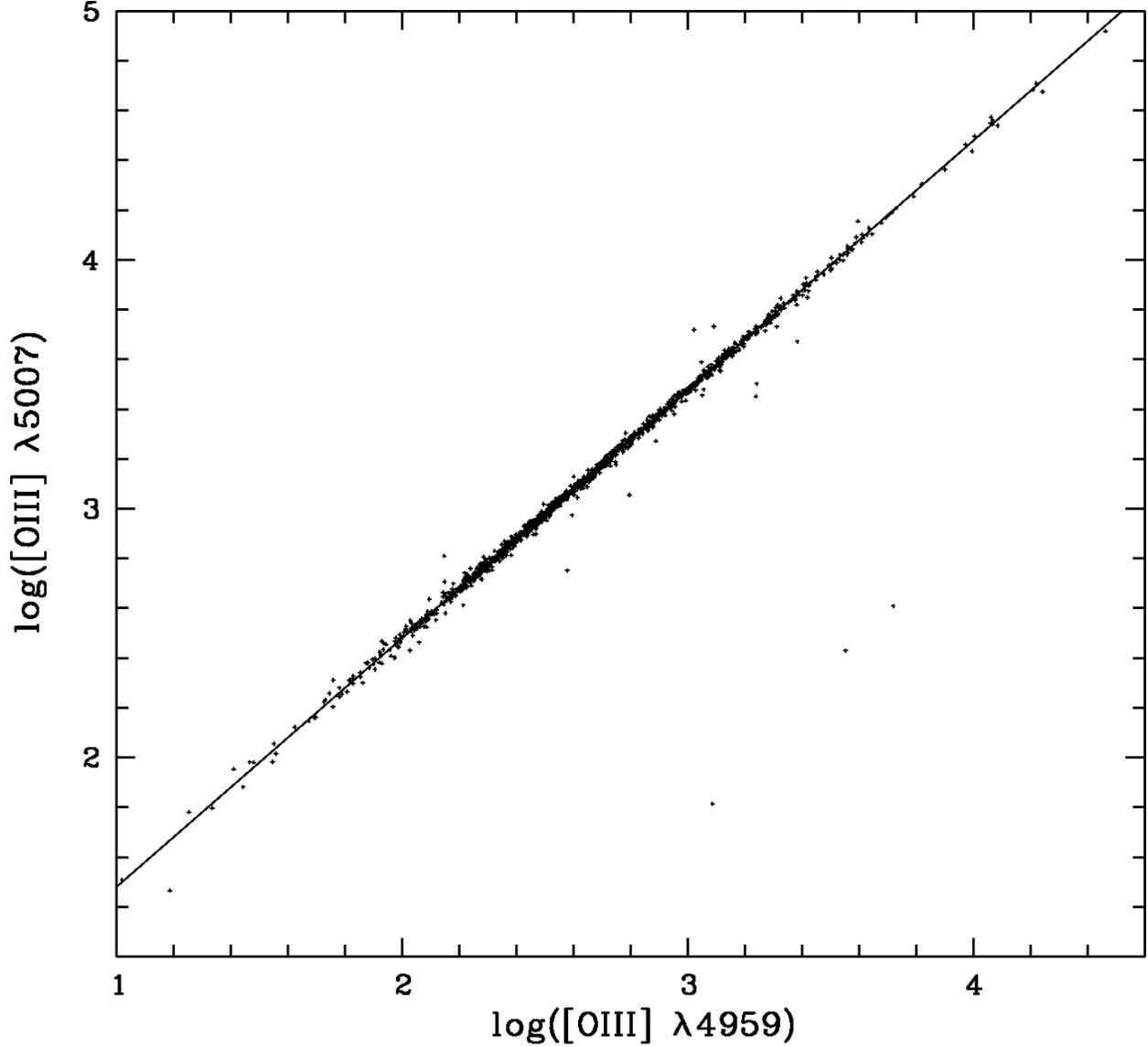}
    \caption{
Plot of the logarithmic flux of [O{\sc iii}] $\lambda$4959 \AA\
versus logarithmic flux of [O{\sc iii}] $\lambda$5007 \AA\
(fluxes for the presented lines are given in units of
10$^{-17}$ erg s$^{-1}$ sm$^{-2}$ \AA$^{-1}$).
The solid line corresponds to the ratio
$F(\lambda5007)/F(\lambda4959) = 3$.
No systematic trend is visible.
Some cases are seen where either the
line [O{\sc iii}] $\lambda$5007 \AA\
(points below the line), or [O{\sc iii}] $\lambda$4959 \AA\
(points above the line) are truncated, which indicates
problems with the SDSS reduction system and/or saturation.
    \label{fig:5007_4959}}
    \end{center}
\end{figure}

The truncation problem was overcome in the process of 
creating the SHOC catalog.
First, in the preselection phase, we extended our selection criteria;
see Section~\ref{txt:sel_steps} for details.
Second, in the process of computing oxygen abundances, most of the 
truncated emission lines were restored, based on ``a priori''
information about line intensity ratios.
In particular, the intensities of the lines [O~{\sc iii}]~$\lambda$5007 and
4959~\AA\ were restored to the adopted ratio of three.
The latter was derived as the mean ratio
$I$([O~{\sc iii}]~$\lambda$5007)/$I$([O~{\sc iii}]~$\lambda$4959)
over all galaxies of the catalog with
unaffected lines [O~{\sc iii}]~$\lambda$5007 and 4959~\AA.
The mean ratio is equal to $3.011\;\pm\;0.017$.
Depending on whether the measured ratio 
[O~{\sc iii}]~$\lambda$5007/[O~{\sc iii}]~$\lambda$4959 
was higher or lower than three, the intensity of either of the two
lines was corrected. 
As is evident in Fig.~\ref{fig:5007_4959},
either line can appear truncated in the SDSS spectral data.
Of course, it is difficult to exclude the possibility that both
the [O\,{\sc iii}]~$\lambda$5007 and 4959~\AA\ lines were truncated.
For this reason, we have indicated objects with truncated lines in
Column 7 of Table~\ref{tab:General}.
But in reality we found only one such spectrum
SDSS J081447.52$+$490400.8 where both
the [O\,{\sc iii}]~$\lambda$5007 and 4959~\AA\ lines were truncated
(see notes to Table~\ref{tab:General} for more details).

\begin{figure}
    \begin{center}
    \epsscale{1.0}
\plotone{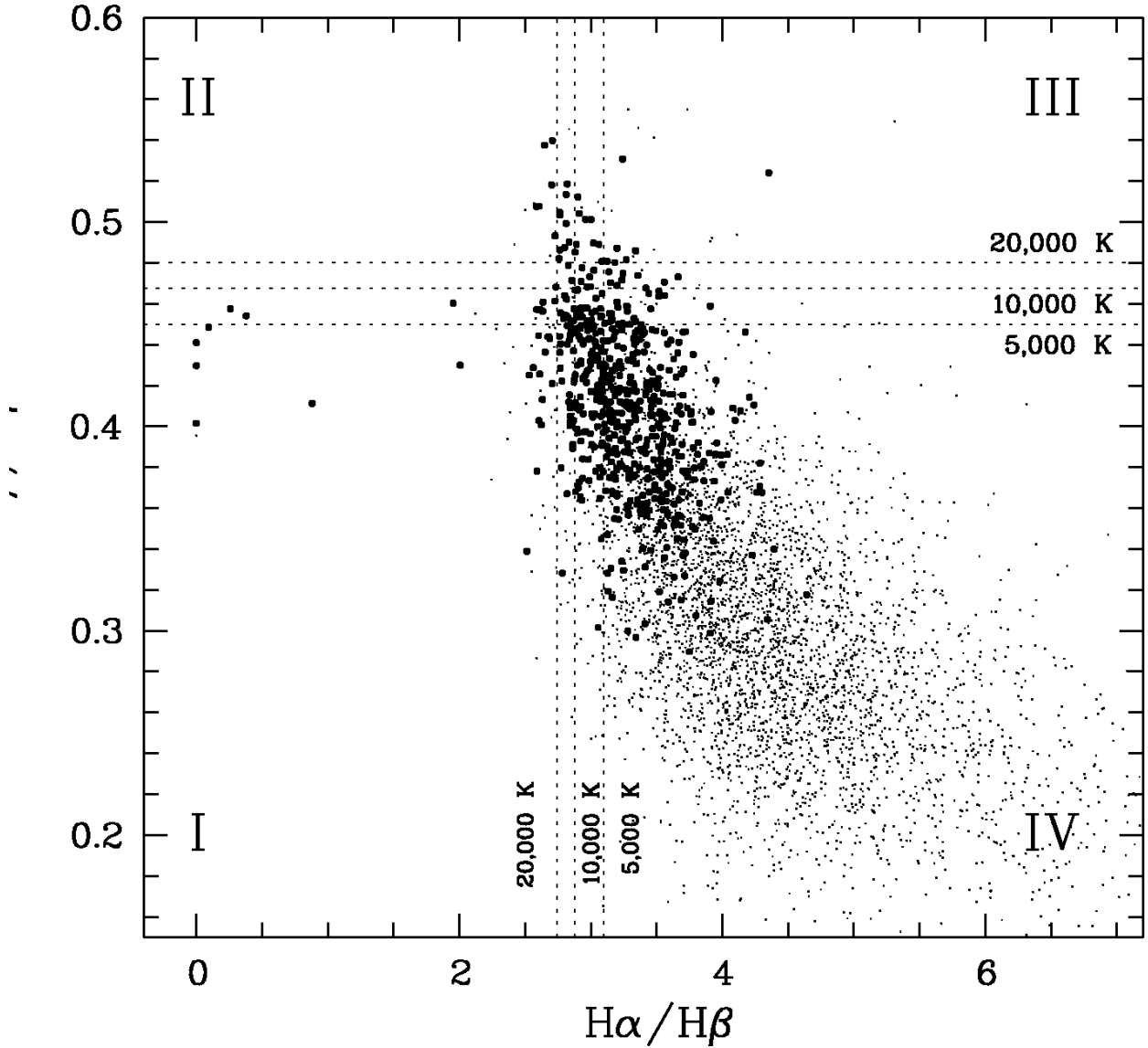}
    \caption{
Plot of the observed
H$\gamma$/H$\beta$ versus H$\alpha$/H$\beta$ flux ratios
to show truncation problems and distribution of these strong line ratios.
The three dotted lines correspond to the theoretical Balmer lines ratios
for T$_\mathrm{e}$ = 5,000, 10,000 and 20,000 K. Filled circles denote
the spectra from the SHOC, while points indicate all other
$\sim$ 5000 preselected ELGs.
Only the area marked ``IV'' is the correct region for the two
Balmer line intensity ratios.
It should be noted that among spectra located in regions
``I'',``II'', and ``III'', truncated spectra are selected
for which the measured line ratios differ from theoretical Balmer
lines ratios by more than 5$\sigma$ 
(see equation \ref{equ:truncation}).
    \label{fig:HaHbHg}}
    \end{center}
\end{figure}

Spectra with truncated H$\alpha$ and/or H$\beta$ emission were
more difficult to treat.
The relation between $I$(H$\alpha$)/$I$(H$\beta$) and
$I$(H$\gamma$)/$I$(H$\beta$) is shown in Fig.~\ref{fig:HaHbHg}.
Only region ``IV'' is the correct region for the two
Balmer line intensity ratios, whose theoretical values for
$T_\mathrm{e}$ = 5,000, 10,000 and 20,000 K are shown by horizontal 
and vertical lines. 
Points located in regions ``I'' and ``II'' of the plot
correspond to the truncation of the H$\alpha$ line.
Points located in regions ``II'' and ``III'' of the plot
correspond to the truncation of the H$\beta$ line.
In region ``II'' of the plot, both H$\alpha$ and H$\beta$ are
truncated.
The H$\beta$ line truncation is most severe for our procedure of the
abundance calculation (see Section~\ref{method_abun} for details), since
all relative intensities I(line)/I(H$\beta$) are affected.
The truncation of the H$\alpha$ line will affect only the calculation
of the extinction coefficient.  However, since this coefficient is
calculated by using all detected Balmer lines (up to eight), the
overall effect of H$\alpha$ line truncation can be rather small.

\subsection{Atmospheric dispersion effects}
\label{txt:atm_disp}

The SDSS does not use an atmospheric refraction corrector, so
the effective fiber position on the sky shifts slightly as a
function of wavelength resulting from atmospheric dispersion
\citep{Fil82}. 
Fixed 3\arcsec\ apertures and the presence of brightness gradients for  
galaxies create errors in measured fluxes.
This is corrected in the SDSS by referring broadband spectrophotometry
to $5.5''\times9''$ aperture ``smear'' exposures \citep{DR1}.
These problems are also reduced for compact objects similar to
H\,{\sc ii} regions in our target galaxies.
As well, the SDSS spectra presented here were taken at airmasses
of 1.22 with r.m.s. 0.10, which further reduces dispersion effects.
If the effect of atmospheric dispersion on our O/H data
remained, we would expect increased data scatter
with larger airmass.
However, our data do not show any trend in the dependence of the error
in oxygen abundance against the airmass of the observed galaxy. 
Finally, we obtain excellent agreement between our measured parameters
and previously published values for already known H{\sc ii} galaxies
recovered from the SDSS data (see Section~\ref{txt:well_known} below).
From this we conclude that the combination of relatively low
airmass and applied `smear' corrections help to reduce significantly
the atmospheric dispersion effect for the sample. The limit is
presumably the value that we currently can check via
comparison of SDSS derived O/H and measurements
from high S/N data in the literature.

\section{Quality of the oxygen abundance determination}
\label{Tests_OH}

Comparing the quality of the oxygen abundance determination for SDSS
ELGs with oxygen abundances from T$_e$ measurements was first done by
\citet{Kniazev03a}, but only for oxygen abundances in the range 7.10
$\le$ 12+log(O/H) $\le$ 7.65.
With our data it is possible to extend the analysis over a much larger
abundance range.

\subsection{Repeatibility}
\label{txt:repeat}

\begin{figure}
    \begin{center}
    \epsscale{0.8}
\plotone{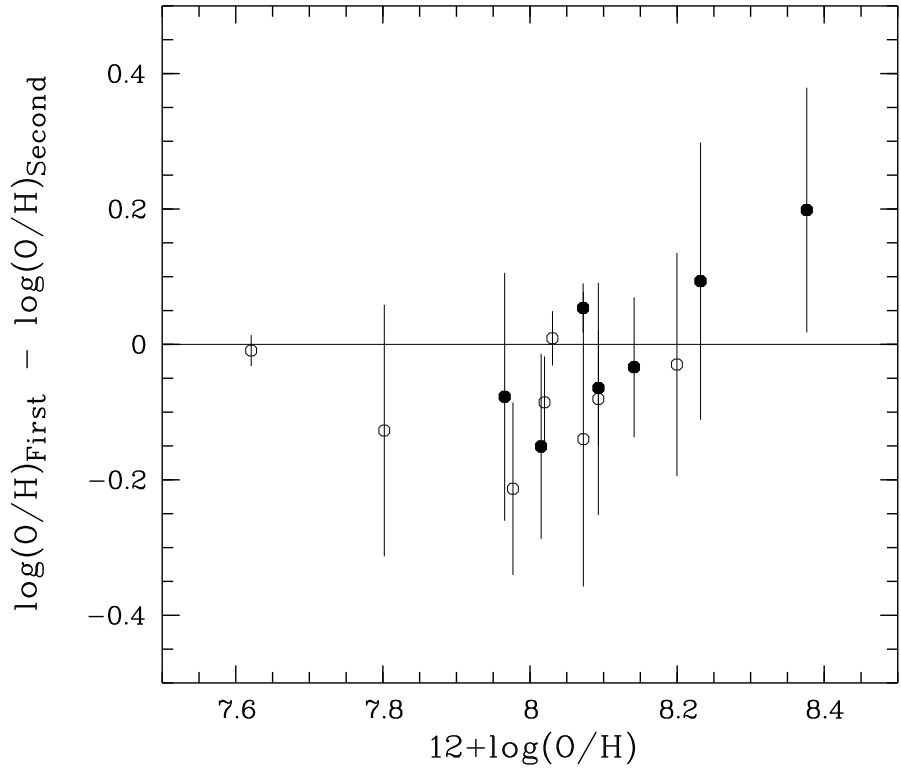}
    \caption{
Comparison of calculated oxygen abundances for catalog galaxies
with two independent observations.
The differences $\Delta$~(First$-$Second) =
$\log$(O/H)$_{\rm First}$--$\log$(O/H)$_{\rm Second}$,
with their total r.m.s.
uncertainties are plotted versus the value 12+log(O/H) for the more accurate
(`First') of two measurements.
The filled circles denote 7 galaxies without detected [O{\sc ii}]
$\lambda$3727 \AA; the [O\,{\sc ii}] $\lambda\lambda$7320,7330 \AA\
lines were used to compute O$^+$/H$^+$.
    \label{fig:repeat_OH}}
    \end{center}
\end{figure}

The comparison of several independent SDSS measurements for the same
object allows a check of the repeatability of our method.
Two independent measurements were available for 14 ELGs.
The results of the comparison are shown in Fig.~\ref{fig:repeat_OH}, where
the differences $\Delta$~(First$-$Second) are plotted versus the value of
12+$\log$(O/H) with the lower r.m.s. uncertainty. The latter we name as the
`First'.
Where independent measurements exist for a given object, the more
accurate value of the oxygen abundance is shown in the catalog.
The shown error bars correspond to the errors of the
differences, taken as a square root of the sum of the errors squared.
The weighted mean of the difference $\Delta$~(First$-$Second) is
$-0.01\pm$0.01 dex with r.m.s. 0.05 dex.
For seven of the 14 objects, [O{\sc ii}] $\lambda$3727 \AA\
was not detected and [O\,{\sc ii}] $\lambda$7320,7330 \AA\ were
used to compute O$^+$/H$^+$.
These objects are shown with filled circles 
in Fig.~\ref{fig:repeat_OH}.
For the seven objects with no [O{\sc ii}] $\lambda$3727 \AA,
the weighted mean of the difference $\Delta$~(First$-$Second) is
0.02$\pm$0.03 dex with r.m.s. 0.07 dex.
These comparisons show that for all repeated observations the oxygen
abundances agree to within the cited uncertainties.

\subsection{Oxygen Abundances with [O~{\sc ii}] 3727,3729~\AA\ 
 and [O~{\sc ii}]  7320,7330~\AA}

\begin{figure}
    \begin{center}
    \epsscale{0.8}
\plotone{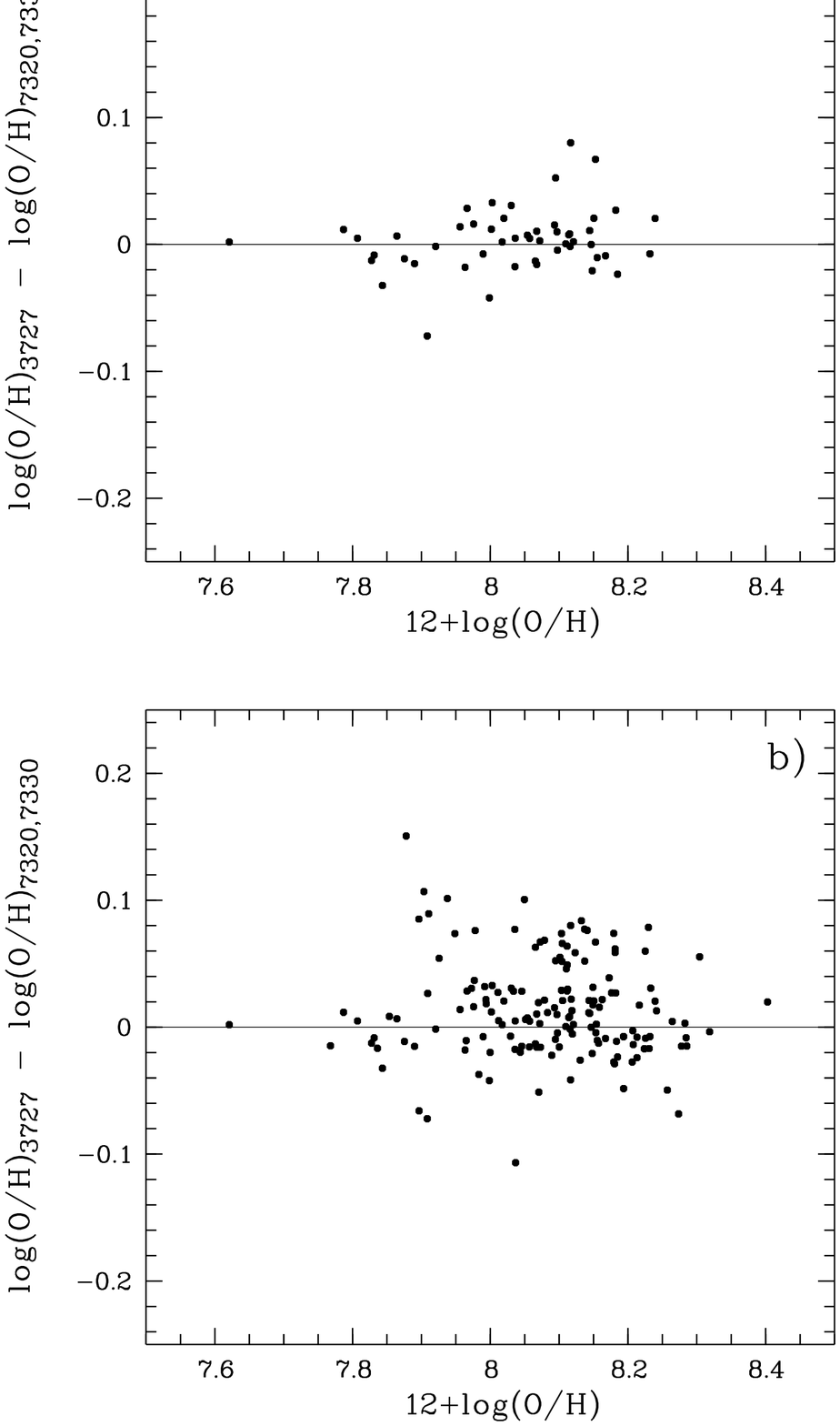}
    \caption{
Differences of logarithmic total oxygen abundances
$\log$(O/H)$_{\rm 3727}$ -- $\log$(O/H)$_{\rm 7320,7330}$
(where $\log$(O/H)$_{\rm 3727}$ means 12+$\log$(O/H)
with the use of [O{\sc ii}] $\lambda$3727 \AA,
and $\log$(O/H)$_{\rm 7320,7330}$ means 12+$\log$(O/H)
with use of [O\,{\sc ii}] $\lambda$7320,7330 \AA\ lines)
versus 12+$\log$(O/H), for objects with $z \ge$~0.024:
a) SHOC galaxies with an accuracy of $\log$(O/H) $\le$ 0.05 dex and
b) SHOC galaxies with an accuracy of $\log$(O/H) $\le$ 0.1 dex.
    \label{fig:3727_7320}}
    \end{center}
\end{figure}

Another type of check can be performed based solely on SDSS spectra.
For many objects with redshifts $\ge$~0.024, the ionic O$^+$ abundance
can be derived from [O{\sc ii}] $\lambda$3727 \AA\ and
[O\,{\sc ii}] $\lambda$7320,7330 \AA\ lines.
A comparison for 51 galaxies with an accuracy of
the oxygen abundance $\le$~0.05 dex and for 159 galaxies
with an accuracy of the oxygen abundance $\le$~0.1 dex
is shown in Fig.~\ref{fig:3727_7320}.
The weighted mean of the difference
$\log$(O/H)$_{\rm 3727}$ -- $\log$(O/H)$_{\rm 7320,7330}$
was found to be 0.002$\pm$0.002 dex with r.m.s. 0.02 dex in both cases,
where $\log$(O/H)$_{\rm 3727}$ is the total oxygen abundance derived using
[O{\sc ii}] $\lambda$3727 \AA,
and $\log$(O/H)$_{\rm 7320,7330}$ is the total oxygen abundance
derived using [O\,{\sc ii}] $\lambda$7320,7330 \AA.
We concluded finally that the scatter between abundances derived with
the [O{\sc ii}] $\lambda$3727 \AA\ and
with the [O\,{\sc ii}] $\lambda$7320,7330 \AA\ lines
is within the cited uncertainties as expected from \citet{Aller84}.

\subsection{The SDSS abundances  versus
	   the data from the literature }
\label{txt:well_known}

\begin{figure}
    \begin{center}
    \epsscale{0.8}
\plotone{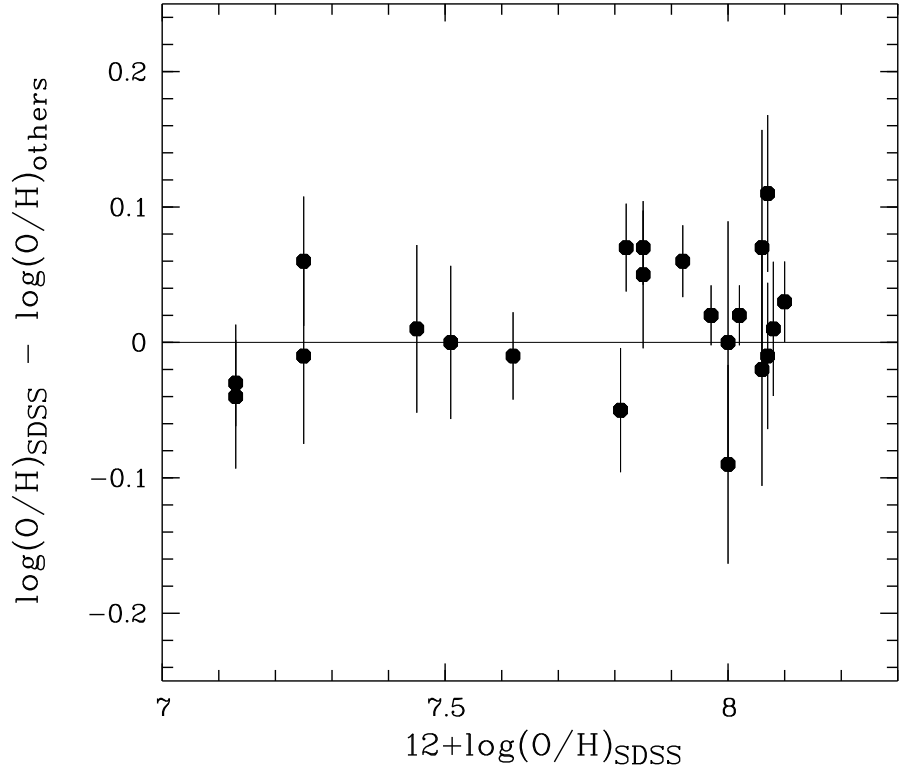}
    \caption{
A comparison of 22 independent measurements of 15 strong-line ELGs
taken from previous works.
Plot of differences of logarithmic total oxygen abundances
$\Delta$~$\log$(O/H) = $\log$(O/H)$_{\rm SDSS}-$$\log$(O/H)$_{\rm others}$
versus 12+$\log$(O/H)$_{\rm SDSS}$,  with the error bars corresponding to
the total r.m.s. uncertainties of these differences.
All presented data are from Table~\ref{tbl:others}.
    \label{fig:reobserved}}
    \end{center}
\end{figure}

A fraction of the strong-line ELGs presented in the SHOC catalog 
is found also in earlier surveys, e.g.,
UM, First Byurakan Survey (FBS or Markarian galaxies -- MRK), SBS,
and HSS-ELG.
For galaxies with sufficiently strong [O{\sc iii}]
$\lambda$4363~\AA\ emission, independent spectrophotometry and oxygen
abundances are published (see notes to Table~\ref{tbl:others} for
references).
These abundances can be used to estimate how reliable our
own abundance determinations are.
A comparison of 22 independent measurements of 15 strong-line ELGs
is shown in Table~\ref{tbl:others}. 
We added also some galaxies from \citet{Kniazev03a} since their
abundances were calculated with the same method.
For all but one of the SDSS spectra, oxygen abundances were calculated
using the intensities of [O{\sc ii}]~$\lambda$7320,7330~\AA\ lines.
We show these abundances in column 6 along with the information
about truncated lines, restored for the abundance calculations.
The weighted mean difference in $\log$(O/H) is 0.02$\pm$0.01 dex with
r.m.s 0.04 dex.
We found that our oxygen abundances derived with
the [O{\sc ii}]~$\lambda$7320,7330~\AA\ lines are consistent within the
cited uncertainties with published abundances in the literature 
derived with [O{\sc ii}]~$\lambda$3727~\AA\ only.
The results are illustrated in Fig.~\ref{fig:reobserved}.
Accounting for probable differences in the centers and sizes
of the regions sampled by SDSS spectra and those used for
comparison, the differences found in oxygen abundances are
satisfactorily small.

\section{Results}
\label{Results}

\subsection{Description of Catalog}

\begin{figure}
    \begin{center}
    \epsscale{0.8}
\plotone{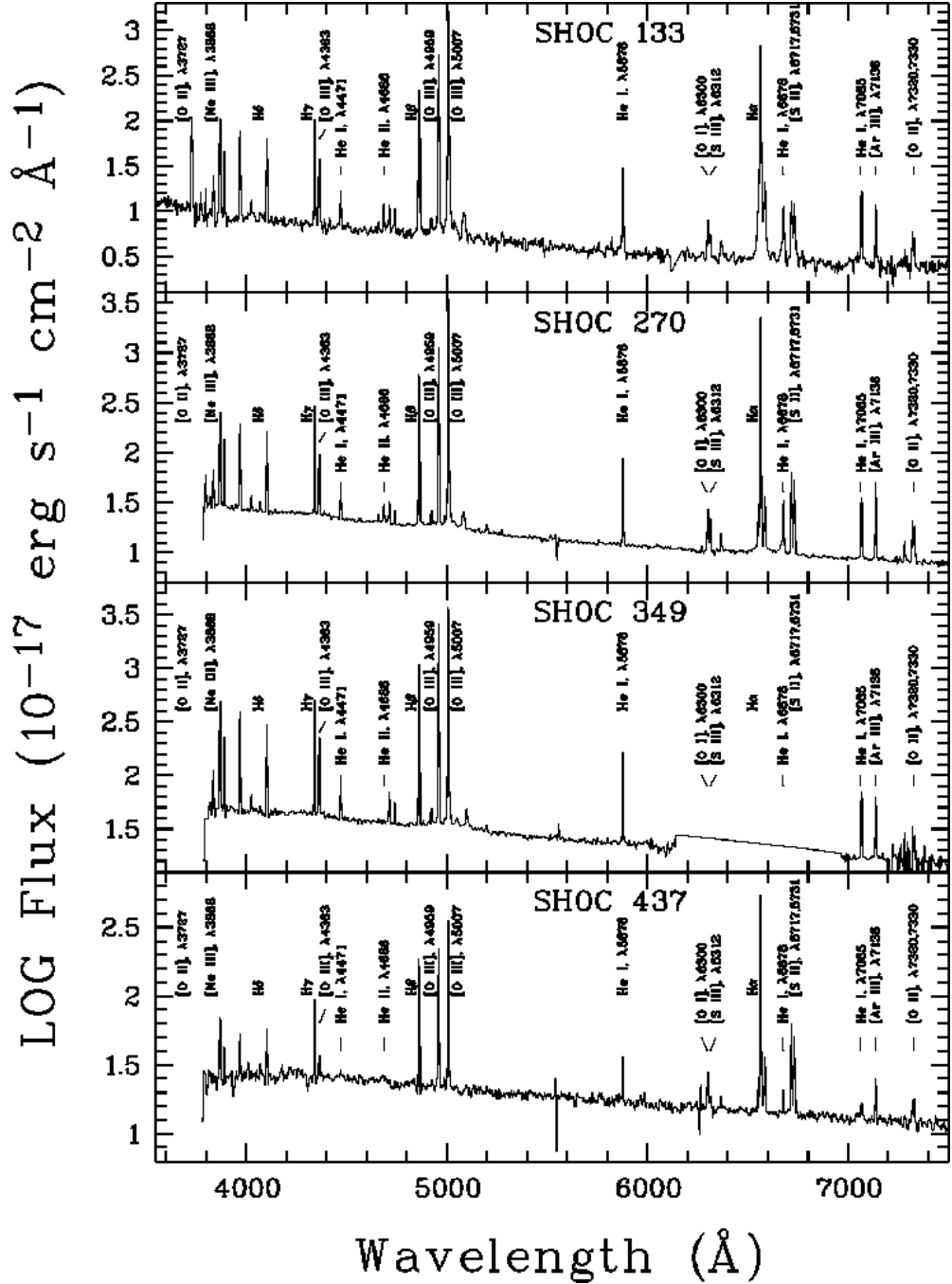}
    \caption{
Examples of spectra from the catalog over the rest-frame
wavelength range of 3600 \AA\ to 7500 \AA. All spectra are shown in
logarithmic scale to see both strong and weak emission lines simultaneously.
Note that of the four galaxies,  the line [O{\sc ii}] $\lambda$3727 \AA\ is
seen only for SHOC~133 (SDSS J024052.20$-$082827.4) (z=0.08221).
In the spectrum of SHOC~349 (SDSS J115133.36$-$022222.0) all lines are truncated
in the spectral region around the H$\alpha$ line.
    \label{fig:Spec_example}}
    \end{center}
\end{figure}

The main criterion for including a galaxy into the SHOC catalog was a 
small r.m.s. uncertainty of the oxygen abundance. 
We set the threshold r.m.s. value of 0.20 dex, corresponding to
$\sim$58\% of the oxygen abundance.
This yielded 612 objects in the catalog with a number of objects
with multiple H{\sc ii} regions (624 separate SDSS targets in total).
263 SHOC objects (272 separate SDSS targets)
have uncertainties $\le$0.10 dex, while 186 objects 
(198 separate SDSS targets)
have intermediate uncertainties between 0.10 and 0.15 dex.
In other words, ELGs with oxygen abundances having uncertainties $\le$0.15
dex make up 75\% of the entire catalog.
In Fig.~\ref{fig:Spec_example} we show representative spectra
of four SHOC objects.

In Table~\ref{tab:General} we present the SDSS parameters of all Catalog
galaxies in the following format.
{\it Column 1}\, lists internal numbers of objects in the catalog.  The
suffixes ``a'', ``b'', and so on represent different star-forming
knots in the same galaxy (sorted by RA).
Entries without an internal number correspond to
a repeat spectrum  of the same object.
{\it Column 2}\, lists the SDSS name according to the coordinates
of the spectral observations and how they exist in the DR1 database.
{\it Column 3}\, lists the plate number,
MJD (Mean Julian Date), and fiber number for spectral observations.
{\it Column 4}\, lists the SDSS $r$-band Petrosian magnitudes.
In many cases these magnitudes relate to the observed H~II regions,
but not to the whole galaxy
due to problems in the pipeline which identify incorrectly
an extended galaxy with a multiple number of some knots or
H~II regions \citep[shredding, see, e.g.,][]{DR1,Kniazev03b}.
We strongly recommend using these cited magnitudes
as preliminary measures.  For further analysis we plan to check
them with the SDSS database or/and with the photometry software of
\citet{Kniazev03b}.
{\it Column 5}\, lists the derived galaxy redshift.
{\it Column 6}\, is the Wolf-Rayet (WR) galaxy flag.  ``1'', ``2'', or
``3'' mean that either only the ``blue'' bump, or only the ``red'' bump,
or both bumps are detected.  Nine previously known or suspected WR
galaxies are shown as asterisks or crosses, respectively.
{\it Column 7}\, lists the Truncation flag.  If ``1'' is in any 
of the four positions,
this indicates that the intensity of H$\beta$, $\lambda$4959,
$\lambda$5007 or H$\alpha$, respectively, was corrected; see 
Section~\ref{Truncation} for details.
{\it Column 8}\, lists the morphological class, as assigned by the
authors, with consultation with the NED database; see details in
Section \ref{Morphology}. 
{\it Column 9}\, lists alternative names for each galaxy, if available,
according to NED. Only names from large surveys for AGN and actively
star-forming galaxies were taken, as well as names from the catalogs
of bright galaxies (UGC, NGC and CGCG).

Observed emission-line fluxes relative to the H$\beta$ emission line,
the H$\beta$ flux, and the H$\beta$ equivalent width in emission
are presented in Table~\ref{t:Intens}. 
In the current work we present only those emission lines,
that were used for oxygen abundance calculations.
Relative emission-line intensities $I(\lambda)$ corrected for
interstellar extinction and underlying stellar absorption 
are presented in Table~\ref{t:Intens_corr}.
The calculated absorption Balmer hydrogen lines equivalent
widths $EW$(abs) and the extinction coefficient $C$(H$\beta$)
are also shown in Table~\ref{t:Intens_corr}.
Derived oxygen abundances and their errors
are presented in Table~\ref{t:Abun}, together with
calculated electron temperatures 
$T_\mathrm{e}$(O~[{\sc iii}]), $T_\mathrm{e}$([O~{\sc ii}]),
electron density $N_\mathrm{e}$([S~{\sc ii}]),
and the ionic abundances O$^{+}$/H$^{+}$ and O$^{++}$/H$^{+}$.
O$^{+}$/H$^{+}$ with [their] cited errors were calculated
using [O\,{\sc ii}] $\lambda$7320,7330 \AA\ lines only
if the line [O{\sc ii}] $\lambda$3727 \AA\ was not detected in
the spectra.

\subsection{Morphology and classification}
\label{Morphology}

The SHOC catalog contains several large groups of gas-rich
H{\sc ii} galaxies and superassociations in the different type of galaxies.
To assign SHOC galaxies to some of the morphological types we examined
combined $g,r,i$ images from SDSS DR1
and produced visual morphological classifications of
five different types:
(1) blue compact galaxies (BCGs) -- if the luminosity of the bright
SF region comprises $\gtrsim$50\% of the total galaxy light;
(2) irregular and/or dwarf (anemic) spirals (Irr) -- if
the luminosity of bright SF knots is less than half of the total
galaxy light; 
(3) low surface brightness galaxies (LSBGs) -- if
the SF region is of very low luminosity;
(4) interacting galaxies with a range of separations (Int);
(5) various types of spiral galaxies (Sp).
In many cases, off- or near-center supergiant H{\sc ii} regions
(super-associations, ``sa'') were clearly observed.
This information was added into our classification scheme.
In intermediate cases and/or in cases which were unclear,
a question mark (``?'') was used as a label.
A significant number of H{\sc ii} galaxies  are found at
large radial velocities/distances ($>$ 20000 km~s$^{-1}$)
and are poorly resolved.
While most of them currently are classified as BCG/BCG?,
their classification requires better angular resolution.
BCGs are the most common type, making up about 75\% (461 galaxies)
of the catalog.
The catalog contains also 78 Irr, 20 LSBGs, 10 Int and 43 obvious spiral
galaxies.

\subsection{Objects with truncated strong lines}
\label{truncated}

In total we found 64 SDSS targets with strong emission lines,
for which one or more lines were truncated in the processing stage with
the SDSS spectroscopic pipeline.
All of these spectra are marked in column 7 of
Table~\ref{tab:General}.
In five SDSS spectra, the H$\beta$ line was truncated and restored.
In 26 SDSS spectra, the H$\alpha$ line was truncated and restored.
In two SDSS spectra, both H$\beta$ and H$\alpha$ were truncated
and restored. 
In 41 SDSS spectra, either the [O{\sc iii}]
$\lambda\lambda$ 4959 or 5007\AA\ line 
was truncated, which was subsequently restored.
Of course, it is difficult to exclude the possibility that both
the [O\,{\sc iii}]~$\lambda$5007 and 4959~\AA\ lines were truncated
for some of these spectra, but
we found only one object (SHOC~193b) for which intensities of
both [O{\sc iii}] $\lambda\lambda$ 4959 \AA\ and 5007\AA\ were truncated.
For this object, intensities of [O{\sc iii}] $\lambda\lambda$
4959 \AA\ and 5007\AA\ were restored based on data from
\cite{Pustilnik00}. 
It should be noted here that all truncated and
restored spectra were also used for the comparisons shown above to check
the quality of oxygen abundance determinations. For example, one
spectrum used to check repeatibility (Section~\ref{txt:repeat}) was
truncated and restored.
Besides, three more truncated and restored spectra were used in
Section~\ref{txt:well_known} (see Table~\ref{tbl:others} for details).
We performed the analysis similar to that described in detail in
Section~\ref{Tests_OH}, but excluding all truncated spectra.
Since the number of truncated
spectra is very small, all our statistics and conclusions
for that Section were not changed.

\subsection{Narrow Line AGN and LINERs}

As mentioned in Section \ref{txt:lines}, in the framework of the
applied algorithm in order to create a continuum appropriate for the
measurements of narrow lines, we adopted the following parameters:
35~\AA\ smoothing window for Gaussian-smoothing convolution and
30 full iterations for the ``smooth-and-cut'' algorithm.
Due to the selected smoothing window size
we strongly suppressed all emission lines with FWHM
$\gtrsim 15$~\AA.  
Galaxies with broad emission lines (FWHM$\gtrsim 15$~\AA, e.g., Sy1)
were lost.
Therefore, the list of AGNs below is incomplete.

\begin{figure}
    \begin{center}
    \epsscale{1.0}
\plotone{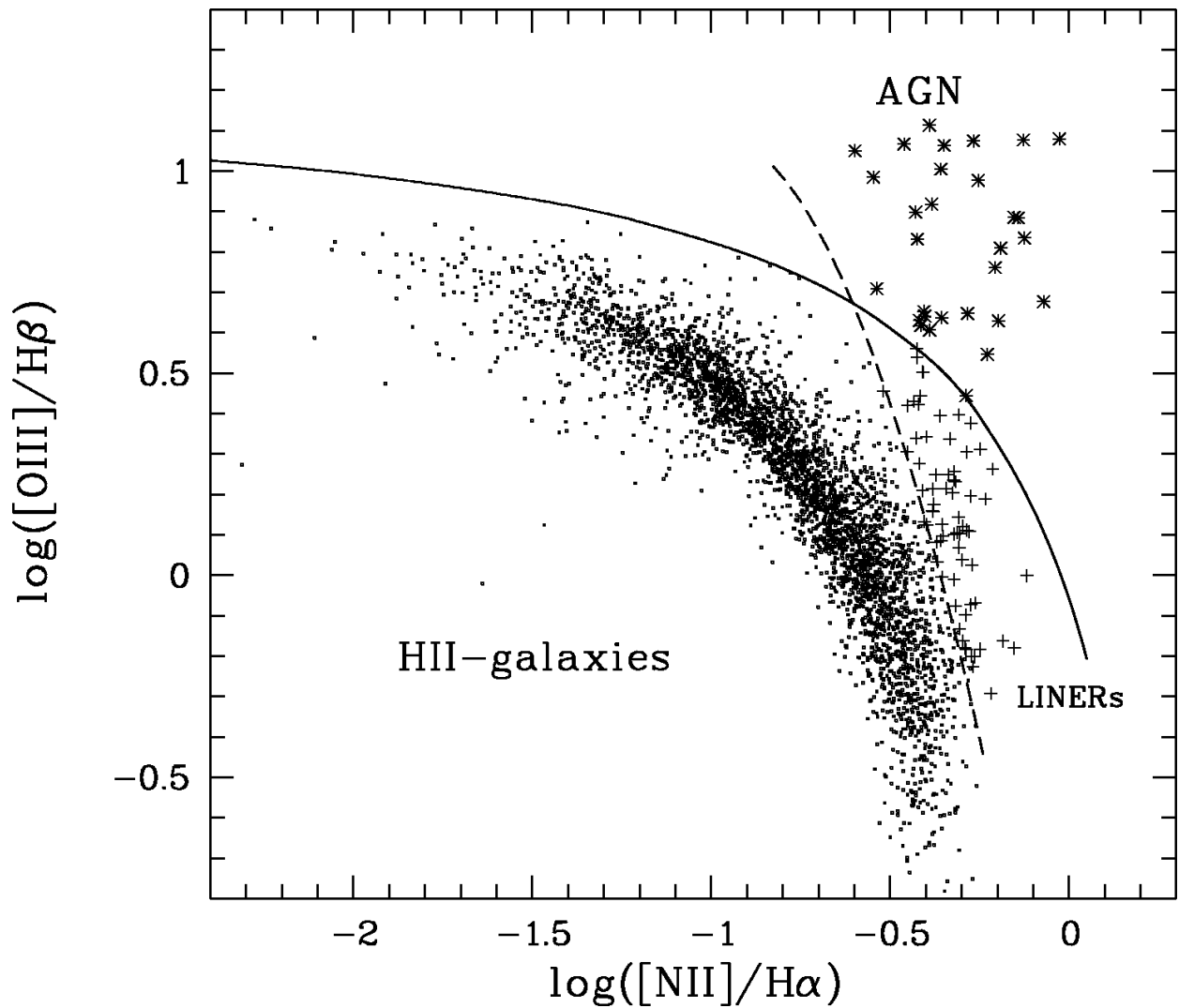}
    \caption{
 Classification diagram for all preselected ELGs ($\sim$~5000 spectra).
 Galaxies identified as AGN are shown as asterisks and
 galaxies identified as LINERs are shown as crosses.
 The other ELGs are plotted as filled circles.
 The dashed line separates
 regions of  HII-type and AGN/LINER spectra following
 \citet{Veilleux87} and  \citet{Baldwin81}.
 The solid line shows models from \citet{Kewley01} that were
 used for AGN/LINER separation.
    \label{fig:class_diag}}
    \end{center}
\end{figure}

In order to distinguish narrow-line AGN and LINERs
from H{\sc ii} galaxies in our sample, we used
classification diagrams based on various line flux ratios,
and models from \citet{Veilleux87} and \citet{Baldwin81}, which
describe the regions occupied by different types of ELGs on these
diagrams.
As an example, we show in Fig.~\ref{fig:class_diag}
the ratios of the line fluxes
[O{\sc iii}]$\lambda$5007/H$\beta$ versus
[N{\sc ii}]$\lambda$6584/H$\alpha$
for $\sim$5000 preselected ELGs and the results of the mentioned models.
This resulted in the identification of 99 galaxies,
which are very probable narrow-line AGN or LINERs.
Most of them have rather large luminosities, corresponding
to $M_{\rm r} \le -20$.
Finally, we used models from \citet{Kewley01}
(the solid line in Fig.~\ref{fig:class_diag}) for
AGN/LINER separation and found 30 narrow-line AGN in our list.
All these AGN and LINERs were not included in the final catalog
and do not have SHOC numbers, but are
presented in a separate list in Table \ref{tab:AGN}.
This includes the SDSS name (column 1), Plate+MJD+Fiber information
for spectral data (column 2), Petrosian $r$ magnitude (column 3), redshift
(column 4), type (column 5),
and comments, which include alternative names, taken from
NED (column 6).

\subsection{WR galaxies}
\label{WR}

Altogether, we detected WR blue ($\sim \lambda$4650~\AA) and/or red
($\sim \lambda$5808~\AA) features in 84 SHOC spectra of 81 SHOC ELGs.
%
%
Twenty-eight additional WR galaxies were found
among galaxies selected at the preliminary step and not included
in the catalog.
They either have an r.m.s. uncertainty of $\log$(O/H) worse than 0.2 dex,
or belong to the AGN/LINER type (4 galaxies).
They are listed in Table~\ref{tab:WR} with the same columns as
in Table~\ref{tab:General}.
These additional WR galaxies do not have SHOC numbers.
The most recent catalog of 139 WR galaxies was compiled
by \citet[][SCP99 hereafter]{Schaerer99}.
Among the WR galaxies found in this work, only
seven  were listed in that catalog and two more were listed by SCP99
as suspected WR galaxies.
Seven WR galaxies from SCP99 are marked in Column~6 of
Table~\ref{tab:General} with an asterisk and 
one of our WR galaxies identified as suspected by SCP99
is marked with a cross in Column~6 of Table~\ref{tab:General}
and one is in Column~6 of Table~\ref{tab:WR}.
The present paper increases the number of known WR galaxies
by $\sim 70$\%.
The most distant known WR galaxy from SCP99 was at
redshift of 0.12. 
The SHOC catalog extends the known population of WR
galaxies out to redshift $z = 0.25$ (SHOC 100).

\subsection{Extremely metal-poor galaxies}
\label{XMP}

Follow-up spectroscopy of the strongest
emisison-line objects from the recent objective-prism based ELG
surveys with limiting $B$-magnitudes of $\sim$18.0--18.5 
(e.g., SBS, HSS-ELG and HSS-LM) indicates that the XMPG surface density
is $3-4$ per 1000 square degrees \citep[e.g.,][]{Kiel02}.
This implies that the SDSS should have another 30--40 BCGs with
$Z < 1/20~Z$\sunn, of which $\sim$2/3 should be new 
objects.
Accounting for SDSS spectroscopy of a significant number of
fainter galaxies, the expected number may even be larger.
The first eight new XMPGs from SDSS
and four rediscovered well-known XMPGs
were identified after an analysis of 250,000 galaxy spectra
within an area of $\sim$3000 deg$^2$ \citep{Kniazev03a}.
All reported XMPGs have uncertainty of
$\log$(O/H) $\le$0.10 dex.

Here, we detected from SDSS DR1 in total 10
galaxies with 12+$\log$(O/H) between 7.25 and 7.65 with
r.m.s. uncertainties below 0.2 dex.  The following four galaxies
have $\log$(O/H) uncertainties lower
than 0.10 dex and were presented in \citet{Kniazev03a}:
SDSS~J020549.13$-$094918.0 (SHOC~106),
HS~0837+4717 (SHOC~220), I~Zw~18 (SHOC~261), and
SDSS~J120122.32$+$021108.3 (SHOC~357).
The remaining galaxies from the catalog
have oxygen abundances near the formal
threshold of 12+$\log$(O/H)=7.65, used to assign objects
to the XMPG group \citep{Kunth2000}.
However, due to the lower accuracy of their abundances, all
these galaxies from the catalog should be checked more carefully.
In particular, one should be aware, that due to the noise
fluctuations, about 5\% of the catalog objects with r.m.s. uncertainties
of O/H $>$0.10 dex can appear to have true abundances differing by
0.20--0.40 dex from 
their catalogued values after higher-quality spectra have been taken.
Such objects may augment the number of currently identified
XMP galaxies.

\section{Discussion}
\label{Discussion}

Our aim in the present work was to create an SDSS based ELG catalog
with reliable oxygen abundances, which would be useful for
various statistical studies related to galaxy chemical
evolution.  
In the current edition of SHOC, we provide only the oxygen abundance.
Argon, neon, nitrogen, and sulphur abundances will be available
for a large number of ELGs in forthcoming papers.

Galaxies selected in the SDSS for spectroscopy come from two
different subsamples.
One is selected by the criterion of the
total $r$-filter magnitude to be brighter than
$r = 17\fm77$ \citep{Strauss02}.
The other fainter subsample is a by-product of a color-selected
sample of quasars \citep{QSO02}.
The SHOC catalog consists of objects selected from both
SDSS galaxy subsamples. The QSO candidate galaxy subsample is,
of course, selected with different criteria
and its study should take this into account.

\begin{figure}
    \begin{center}
    \epsscale{1.0}
\plotone{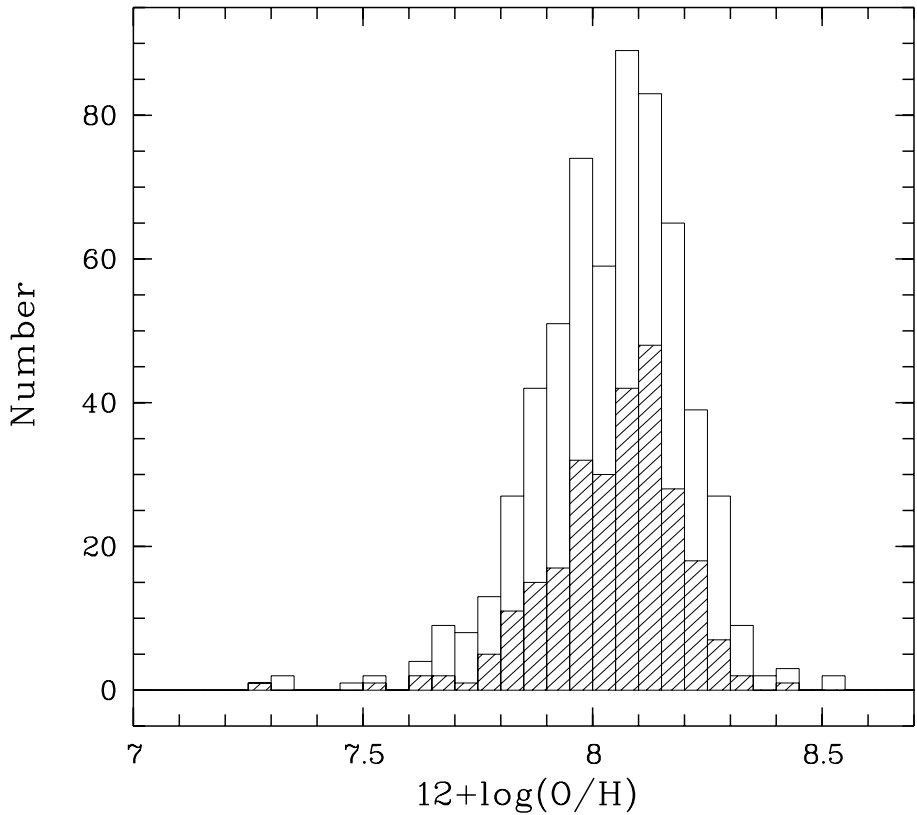}
    \caption{
The metallicity distribution of all ELGs from our catalog,
as measured by their oxygen abundances.
Where observations exist for multiple knots in a given galaxy,
only metallicity for knot ``a'' is shown.
The open histogram shows the metallicity distribution for
612 galaxies with an accuracy of $\log$(O/H) $\le$ 0.2 dex
(mean = 8.04, $\sigma$ = 0.16, median = 8.06).
The hashed histogram shows the metallicity distribution for
263 galaxies with an accuracy of $\log$(O/H) $\le$ 0.1 dex
(mean = 8.05, $\sigma$ = 0.14, median = 8.07).
    \label{fig:OH_all}}
    \end{center}
\end{figure}

\begin{figure}
    \begin{center}
    \epsscale{0.7}
\plotone{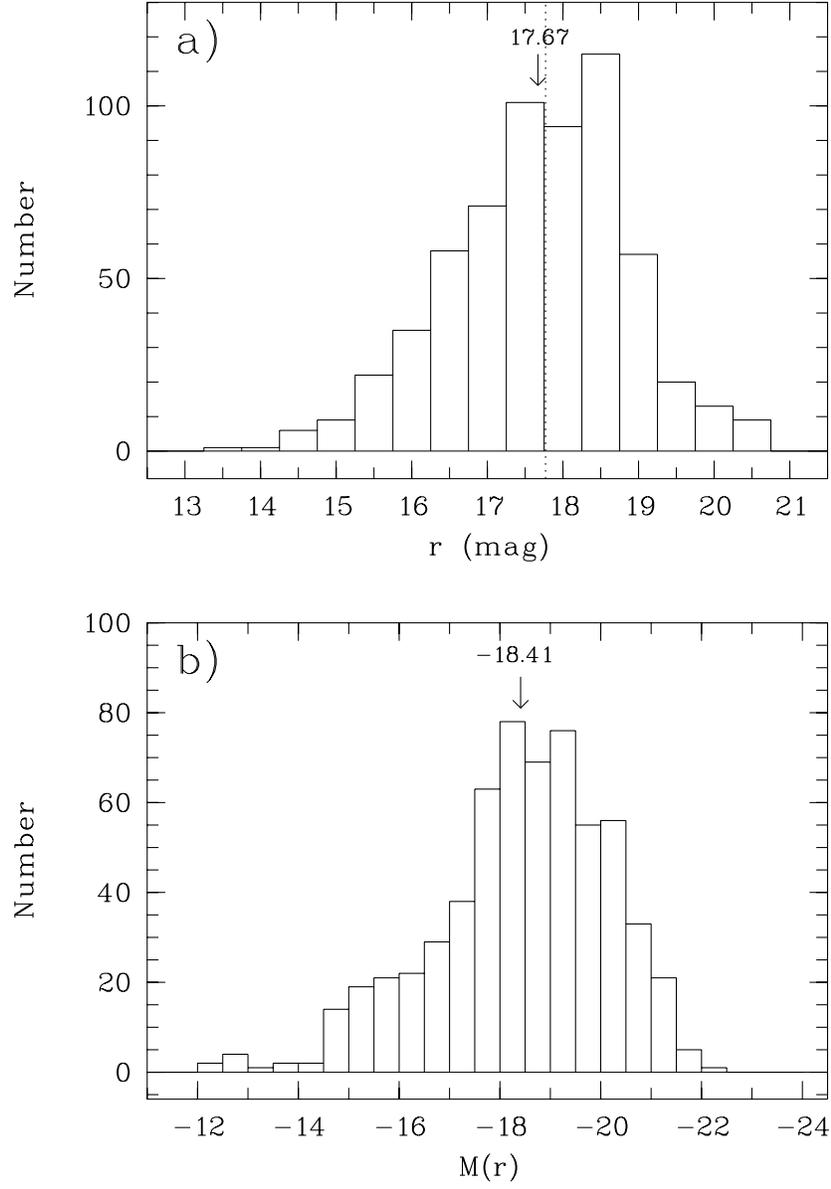}
    \caption{
 Distributions of apparent $r$-magnitudes (a) and absolute
$r$-magnitudes (b)
for SDSS-selected H{\sc ii} galaxies
with accuracies of $\log$(O/H) $\le$ 0.2 dex.
The arrows indicate median values.
The dotted vertical line in panel (a) marks the 
SDSS spectroscopic limit $r$ = 17\fm77 
\citep{Strauss02}.
    \label{fig:Catalog}}
    \end{center}
\end{figure}

\begin{figure}
    \begin{center}
    \epsscale{0.7}
\plotone{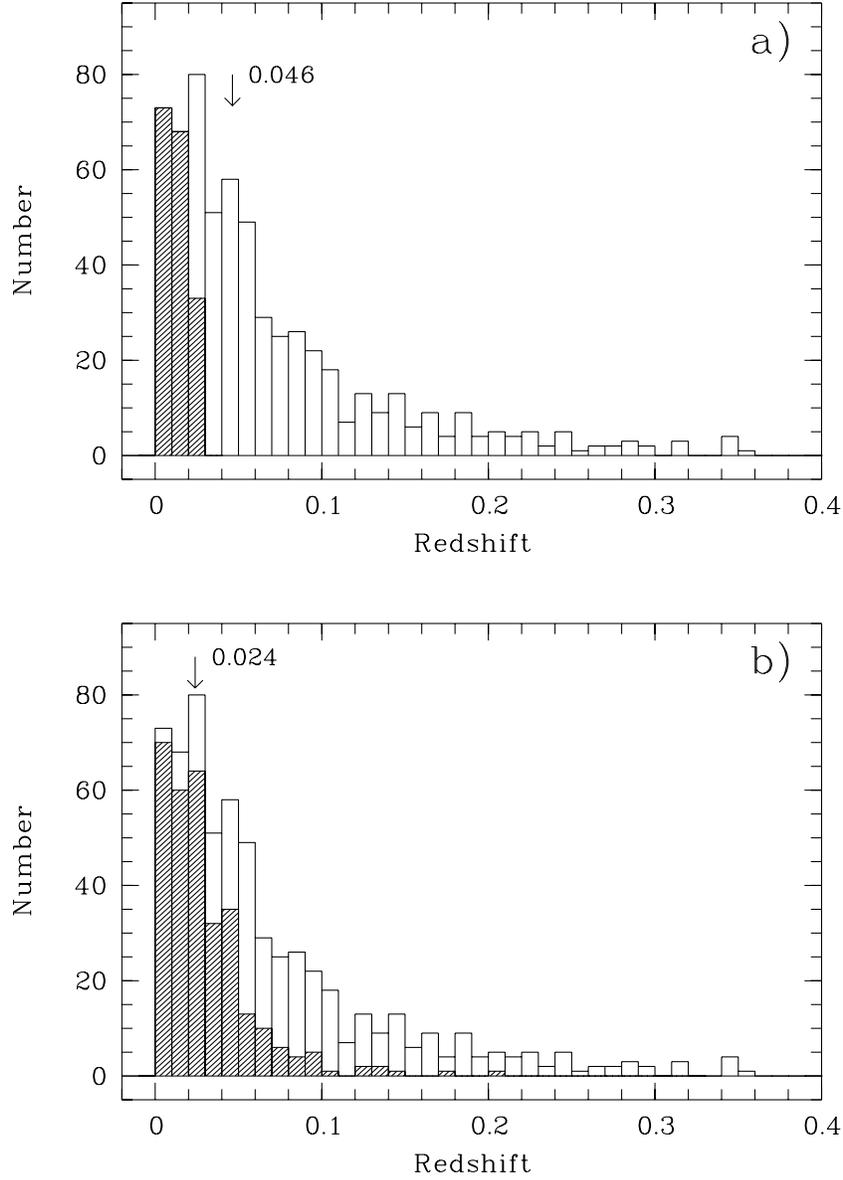}
    \caption{
Distributions of redshifts for the SHOC galaxies.
a) All 612 galaxies (open histogram) and 174 objects
(hashed histogram)
for which only [O\,{\sc ii}] $\lambda$7320,7330 \AA\ were used
for O$^+$/H$^+$ calculation. The arrow indicates the median redshift
for all galaxies.
b) All galaxies (open histogram) and magnitude limited sample of galaxies
with r-magnitudes are brighter than 17\fm77 (hashed histogram).
The arrow 
indicates the median redshift for the magnitude-limited sample.
    \label{fig:redsh}}
    \end{center}
\end{figure}

To demonstrate the range of key galaxy parameters for 
catalog objects, we plot them in Fig.~\ref{fig:OH_all}--~\ref{fig:redsh}.
First, we show in Fig.~\ref{fig:OH_all} the distribution of
12+$\log$(O/H) derived for all 612 SHOC ELGs
(in case of observations for multiple knots in a given
galaxy, only metallicity for knot ``a'' is shown).
The open histogram outlines the distribution of all ELGs with
r.m.s. uncertainties of $\log$(O/H) $\le$0.20 dex.
The hashed histogram shows the same distribution for 
the subsample with $\log$(O/H) uncertainties $\le$0.10 dex.
Both distributions appear similar and their similar mean
values and the standard deviations confirm this impression. 
Not surprisingly, the subsample with larger uncertainties exhibits
somewhat larger scatter.
In Figure~\ref{fig:Catalog} we show the distributions
of apparent $r$ magnitudes, absolute magnitudes $M_r$, and
heliocentric radial velocities for all 612 SHOC ELGs.
The median apparent magnitude is $m_r^{\rm med}$=17\fm67
(top panel on Figure~\ref{fig:Catalog}), which indicates
that more than 50\% of SHOC galaxies are in the magnitude range
``completely'' sampled by SDSS spectroscopy.
The median value of the absolute magnitude $M_r$ is
$M_{r}^{\rm med} = -18\fm4$.
The characteristic $r$-band luminosity $L_r^{\ast}$ for
the SDSS-determined luminosity function (for a Hubble constant
75~\kms~Mpc$^{-1}$) corresponds to $M_r^{\ast} =-21\fm45$
\citep[see, e.g.,][]{Blanton01}.
About 90\% of the Catalog galaxies are subluminous 
($M_r > M_r^{\ast} + 1$).

The number distribution of galaxies with radial velocities
decreases above $V_{hel} \sim$10,000~\kms\ up to
$V_{hel} \sim$110,000~\kms (top panel on Figure~\ref{fig:redsh}).
About 90\% of all ELGs have $V_{hel} \le $55,000~\kms.
ELGs with redshifts $z \le 0.024$, for which
[O{\sc ii}]$\lambda$3727 is not measurable,
are shown by hashed bins on the top panel on Figure~\ref{fig:redsh}.
They comprise $\sim$30\% of the total number
of SHOC galaxies.
The median radial velocity of $\sim$14,000~\kms\
on the top panel on Figure~\ref{fig:redsh}
reflects the presence in the catalog of a significant number of
rather distant galaxies with H{\sc ii} type spectra.
If we limit the subsample of galaxies to those selected for
SDSS spectroscopy with r-magnitudes brighter than 17\fm77,
this subsample is about 50\% of the total number
(hashed histogram on bottom panel of Figure~\ref{fig:redsh}).
The redshift distribution is more narrow for this subsample:
its median radial velocity is $\sim$7000~\kms\ with about 90\%
of all galaxies having $V_{hel} \le $17,000~\kms.

It is interesting to compare this catalog with other samples of
BCG/H{\sc ii} galaxies with O/H measured by the classic $T_{e}$ method.
They include the BCG samples from 
the Second Byurakan Survey (SBS) \citep{IT99},
the Tololo and UM surveys \citep{Mas94},
the KISS \citep{Melbourne02}
and the HSS-LM \citep{Ugryumov03}.

For the SBS BCGs the value of O/H is measured for about 40 objects with
r.m.s. uncertainties of $\log$(O/H) in the range of 0.01 to 0.05
dex \citep{IT99}.
Their published magnitudes are known with moderate to low
precision, so it is quite difficult to address the issue 
of completeness.

For Tololo and UM galaxies \citep{Terlevich91},
oxygen abundances are presented for 100 galaxies \citep{Mas94}.
Their cited r.m.s. uncertainties are between 0.01 and 0.08 dex.
However, checks of several objects from that work
\citep[e.g.,][]{Kniazev01,Pustilnik02a} showed that the real
uncertainties are larger,
with some values shifted by about 0.3 dex.
For a part of this sample of galaxies, the
magnitudes are poorly known.

For the KISS sample the number of ELGs with strong lines is
significantly larger. 
However, only 12 galaxies have been published with oxygen abundances derived with
the $T_{e}$ method \citep{Melbourne02}.
No estimates of their uncertainties are presented.
These ELGs are on average more distant, since KISS
includes galaxies with $m_{B}$ as faint as 20$^m$.

For the HSS-LM published List I \citep{Ugryumov03}
there are 46 strong-lined BCG/H{\sc ii} galaxies 
with $T_{e}$ oxygen abundances; the faintest galaxy is about
$m_{B}$ = 18\fm5.
The uncertainties of $\log$(O/H) for this sample vary between 0.02 and 0.20
dex. List~II (Pustilnik et al., in preparation) will
augment a comparable number of H{\sc ii} galaxies with
similar properties.

\begin{figure}
    \begin{center}
    \epsscale{0.7}
\plotone{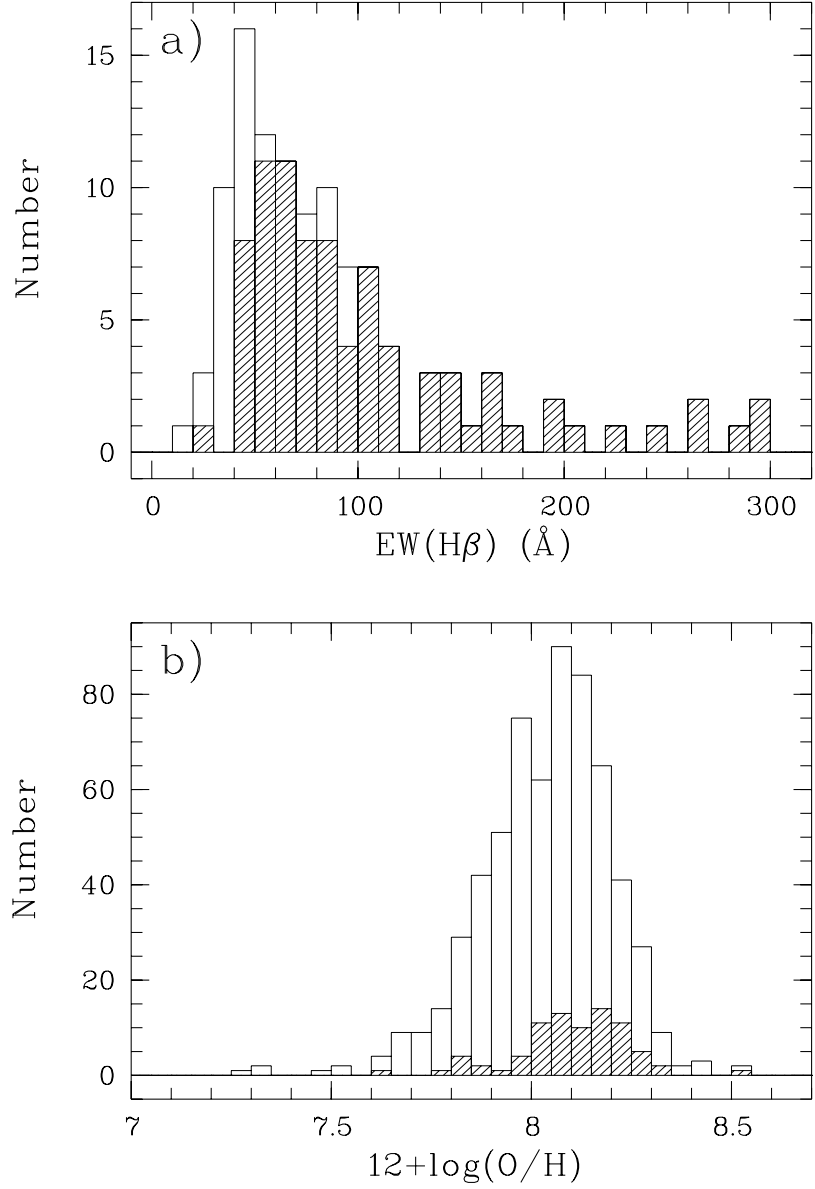}
    \caption{
a) Distribution of EW(H$\beta$) for all 109 detected WR-galaxies
from this work (open histogram) and 81 WR-galaxies
from the catalog (hashed histogram).
b) The oxygen abundance distribution (open histogram) of galaxies
from the SHOC and similar distribution of WR-galaxies (hashed histogram)
in the SHOC (81 galaxies).
    \label{fig:WR}}
    \end{center}
\end{figure}

For the detected WR galaxies we notice that a majority of the
objects have EW(H$\beta$) larger than 60~\AA\ (see Figure~\ref{fig:WR},
top panel),
with the median of 72 \AA\ for all detected WR galaxies and
86 \AA\ for the WR galaxies in the catalog.
This is consistent with expectations from current models of
massive star evolution, which predict the presence of an observable
number of WR stars for starburst ages between 3 and 6 Myr,
for the metallicity range of SHOC objects \citep{SV98}.
The maximum of the WR fraction in the catalog galaxies with
oxygen abundances 12+$\log$(O/H)$\gtrsim$8.1 (Figure~\ref{fig:WR}, bottom
panel) is as well consistent with the model predictions of the dependence
of the WR-bump strength and duration on the galaxy metallicity.

Thus, the SHOC catalog of ELGs with measured oxygen abundances
presents three key advantages.
First, SHOC is by number several times larger than any of the
previously published similar samples. 
Two, almost all SHOC galaxies have potentially good photometry.
However, in many cases derived photometry of SDSS data requires
additional software \citep[e.g., used in][]{Kniazev03b} to avoid
galaxy shredding.
Finally, about 50\% of the SHOC galaxies with $m_r < 17\fm7$ have been
observed spectroscopically to a high level
of completeness \citep[$\sim$0.9;][]{Strauss02}.
The SHOC catalog provides a significant opportunity to improve the
statistical study of many issues related to the metallicity of
gas-rich galaxies.

\section{Conclusions}
\label{Summary}

With respect to the present SHOC catalog
and data from \citet{Kniazev03a},
we draw the following conclusions:
\begin{itemize}
\item
SDSS spectra permit accurate oxygen abundance determinations
over the range $7.1 \la 12+\log({\rm O/H}) \la 8.5$.
\item
The method for calculating O$^+$/H$^+$ using intensities of the
[O\,{\sc ii}] $\lambda$7320,7330 \AA\ lines
appears to yield reliable results over a wide
range of oxygen abundances.
\item
A large number of strong-line ELGs with measurable oxygen abundances
and detectable WR populations is selected from the SDSS DR1 database.
\item
A large majority of strong-line ELGs with detected
[O~{\sc iii}]$\lambda$4363 are H{\sc ii} galaxies with a broad
range of $r$-band luminosities, corresponding to absolute $r$
magnitudes $-22 \la M_r \la -12$.
\end{itemize}

We plan to produce regular updates of the SHOC catalog of strong-line
ELGs with measured oxygen abundances, based on subsequent Data
Releases from the SDSS.

\acknowledgments
The Sloan Digital Sky Survey (SDSS) is a joint project of The University of
Chicago, Fermilab, the
Institute for Advanced Study, the Japan Participation Group, The Johns
Hopkins University, the
Max-Planck-Institute for Astronomy (MPIA), the Max-Planck-Institute for
Astrophysics (MPA),
New Mexico State University, Princeton University, the United States Naval
Observatory, and the
University of Washington. Apache Point Observatory, site of the SDSS
telescopes, is operated by the
Astrophysical Research Consortium (ARC).

Funding for the project has been provided by the Alfred P. Sloan Foundation,
the SDSS member
institutions, the National Aeronautics and Space Administration, the National
Science Foundation, the
U.S. Department of Energy, the Japanese Monbukagakusho, and the Max Planck
Society. The SDSS Web site is http://www.sdss.org/.

This research has made use of
the NASA/IPAC Extragalactic Database (NED) which is operated by the
Jet Propulsion Laboratory, California Institute of Technology, under
contract with the National Aeronautics and Space Administration.

\clearpage

\clearpage

\newcounter{tab}
\setcounter{tab}{0}
\newcommand{\numb}{\addtocounter{tab}{1}\arabic{tab}}

\begin{deluxetable}{lllllllll}
\rotate
\tabletypesize{\scriptsize \bf}
\tablecaption{General parameters of the
	      Strong Emission Line H{\sc ii} galaxies from SDSS DR1
	      with an oxygen abundance accuracy of $\le$0.2 dex
\label{tab:General}
}
\tablewidth{0pt}
\tablehead{
\colhead{SHOC ID\tablenotemark{a}}    & \colhead{SDSS name} & \colhead{Plate,MJD,Fiber} &
\colhead{$r$}   & \colhead{$z$}       &     \colhead{$WR$\tablenotemark{b}}      &
\colhead{Trun} & \colhead{Class} & \colhead{Comments}  \\
\colhead{(1)} & \colhead{(2)} & \colhead{(3)} &
\colhead{(4)} & \colhead{(5)} & \colhead{(6)} & \colhead{(7)} &
\colhead{(8)} & \colhead{(9)}
}
\startdata
1                     & SDSS J000439.33$-$100909.3 &   650 52143  61 & 17.93 &  0.10842 & ~0~ & 0000 & BCG     &                                                               \\
2                     & SDSS J000657.03$+$005125.9 &   388 51793 457 & 18.81 &  0.07358 & ~0~ & 0000 & BCG     &     UM 199                                                    \\
3                     & SDSS J000703.99$-$003447.5 &   388 51793 239 & 18.69 &  0.12129 & ~0~ & 0000 & BCG     &                                                               \\
4                     & SDSS J000706.61$-$085051.1 &   651 52141 408 & 18.04 &  0.03568 & ~0~ & 0000 & BCG     &                                                               \\
5                     & SDSS J000847.56$-$085905.5 &   651 52141 452 & 17.62 &  0.02662 & ~0~ & 0000 & BCG     &                                                               \\
6                     & SDSS J001327.48$-$001855.9 &   389 51795 266 & 18.54 &  0.04329 & ~0~ & 0000 & BCG     &     UM 214                                                    \\
7                     & SDSS J001354.80$+$005427.3 &   389 51795 405 & 19.16 &  0.08466 & ~0~ & 0000 & BCG     &                                                               \\
8                     & SDSS J001647.76$-$104742.3 &   652 52138  90 & 16.08 &  0.02324 & ~0~ & 0000 & Irr?    &                                                               \\
\enddata
\tablenotetext{a}{The number of individual galaxy in the SHOC catalog.
Letters `a' and `b' indicate separate regions within the same galaxy}
\tablenotetext{b}{The Wolf-Rayet galaxy flag.  ``1'', ``2'', or
``3'' mean that either only the ``blue'' bump, or only the ``red'' bump,
or both bumps are detected. 
Asterisks identify WR galaxies
from SCP99.  Crosses identify WR galaxies from the list of suspected
WR galaxies in SCP99.
} 
\tablenotetext{c}{SHOC~11 was selected as a Sy~2 galaxy (LBQS 0018+0036)
by \citet{L79}, and is assigned as an AGN in
NED. However, from both \citet{Terlevich91} (UM~228) and from SDSS data this
is classified as an H{\sc ii} galaxy.}
\tablenotetext{d}{
  {SHOC~56 is not UM~311 as identified in NED, but is
  another superassociation in the spiral galaxy NGC~0450 as seen
  by comparing SDSS images with those in \citet{MacAlpine78}.}
}
\tablenotetext{e}{SHOC~193b (SDSS J081447.52+490400.8) has very strong
lines. This is a giant H{\sc ii} region on the periphery of a
nearby SA(s)cd galaxy NGC~2541, also known as HS~0811+4913 \citep{Pustilnik00}.
For this object,
both [O{\sc iii}] $\lambda\lambda$ 4959 and 5007\AA\ were truncated,
while H$\alpha$ and H$\beta$ are not.
To correct an oxygen abundance the intensity  ratio
[O{\sc iii}]~$\lambda$~5007~\AA/H$\beta$ = 7.37 was taken from 
\cite{Pustilnik00}.
}
\end{deluxetable}

\begin{deluxetable}{lcrrrrrrrrrrrcrr}
\rotate
\setlength{\tabcolsep}{0.02in}
\tabletypesize{\tiny \bf}
\tablecaption{Observed Relative Emission Line Fluxes and Errors\tablenotemark{a}.
\label{t:Intens}
}
\tablewidth{0pt}
\tablehead{
\colhead{SHOC\tablenotemark{b}} & \colhead{[O\,{\sc ii}]}& \colhead{H10}       &\colhead{H9}         &\colhead{H$\delta$} & \colhead{H$\gamma$} & \colhead{[O\,{\sc iii}]} & \colhead{H$\beta$}  & \colhead{[O\,{\sc iii}]} & \colhead{[O\,{\sc iii}]} & \colhead{H$\alpha$} & \colhead{[S\,{\sc ii}]} & \colhead{[S\,{\sc ii}]} & \colhead{[O\,{\sc iii}]} & \colhead{W(H$\beta$)\tablenotemark{c}} & \colhead{F(H$\beta$)\tablenotemark{d}} \\
\colhead{ID}                    & \colhead{3727 \AA}     &  \colhead{3798 \AA} &  \colhead{3835 \AA} &\colhead{4101 \AA}  &  \colhead{4340 \AA} &  \colhead{4363 \AA}      & \colhead{4861 \AA}  & \colhead{4959 \AA}       & \colhead{5007 \AA}       & \colhead{6563 \AA}  & \colhead{6716 \AA}      & \colhead{6731 \AA}      & \colhead{7320+7330 \AA}  & \colhead{4861 \AA}                     & \colhead{4861 \AA}    \\
\colhead{(1)}                   & \colhead{(2)}          &  \colhead{(3)}      &  \colhead{(4)}      & \colhead{(5)}      & \colhead{(6)}       & \colhead{(7)}            & \colhead{(8)}       & \colhead{(9)}            & \colhead{(10)}           & \colhead{(11)}      & \colhead{(12)}          & \colhead{(13)}          & \colhead{(14)}           & \colhead{(15)}                         & \colhead{(16)}
}
\startdata
1    &  191.3$\pm$ 3.7 &    \nodata   &  1.5$\pm$1.0 & 17.4$\pm$1.3 & 38.2$\pm$1.2 &  2.4$\pm$0.8 & 100.0$\pm$1.3 & 113.0$\pm$ 1.8 & 341.5$\pm$ 4.7 & 428.4$\pm$ 5.8 & 43.8$\pm$ 1.4 & 31.7$\pm$ 1.3 &   6.1$\pm$2.0 &    50.6  &   3.93   \\
2    &  174.7$\pm$ 2.9 &  2.1$\pm$0.7 &  2.5$\pm$0.8 & 21.1$\pm$1.4 & 43.2$\pm$0.7 &  7.2$\pm$0.5 & 100.0$\pm$0.6 & 181.4$\pm$ 1.3 & 556.1$\pm$ 3.6 & 305.3$\pm$ 2.0 & 17.7$\pm$ 0.8 & 13.7$\pm$ 0.8 &   5.1$\pm$1.1 &    87.6  &   9.05   \\
3    &  234.0$\pm$ 5.3 &  1.4$\pm$0.8 &  4.4$\pm$1.1 & 27.7$\pm$1.2 & 48.9$\pm$1.1 &  7.6$\pm$0.9 & 100.0$\pm$1.1 & 188.0$\pm$ 2.6 & 560.3$\pm$ 7.0 & 289.0$\pm$ 3.4 & 17.3$\pm$ 1.0 & 12.8$\pm$ 1.0 &   3.8$\pm$1.1 &    67.4  &   6.32   \\
4    &  278.0$\pm$ 9.1 &     \nodata  &    \nodata   & 14.2$\pm$3.4 & 38.0$\pm$1.9 &  3.9$\pm$1.5 & 100.0$\pm$2.3 &  99.8$\pm$ 2.8 & 304.7$\pm$ 7.3 & 307.0$\pm$ 7.2 & 35.6$\pm$ 1.4 & 24.7$\pm$ 1.1 &   9.7$\pm$2.8 &    22.8  &   4.36   \\
5    &  262.8$\pm$ 6.0 &     \nodata  &    \nodata   & 19.3$\pm$1.6 & 41.7$\pm$1.3 &  3.8$\pm$0.9 & 100.0$\pm$1.3 & 150.7$\pm$ 2.2 & 460.7$\pm$ 6.4 & 304.2$\pm$ 4.1 & 28.5$\pm$ 0.7 & 18.8$\pm$ 0.6 &   5.6$\pm$0.7 &    41.8  &  10.85   \\
6    &  262.1$\pm$ 6.0 &    \nodata   &  2.9$\pm$1.5 & 23.3$\pm$2.3 & 46.7$\pm$1.3 &  1.7$\pm$0.9 & 100.0$\pm$1.3 & 137.1$\pm$ 1.9 & 411.1$\pm$ 5.5 & 288.0$\pm$ 3.8 & 23.3$\pm$ 0.7 & 17.3$\pm$ 0.7 &   2.8$\pm$0.6 &    51.4  &   5.02   \\
7    &  238.5$\pm$ 6.3 &    \nodata   &  3.5$\pm$2.0 & 24.3$\pm$2.9 & 45.8$\pm$2.1 &  3.2$\pm$1.6 & 100.0$\pm$1.6 &  84.4$\pm$ 1.7 & 247.9$\pm$ 4.2 & 345.6$\pm$ 5.9 & 41.1$\pm$ 1.7 & 30.1$\pm$ 1.6 &   2.5$\pm$1.2 &    50.3  &   3.73   \\
8    &       \nodata   &  1.8$\pm$0.7 &  1.7$\pm$0.8 & 16.8$\pm$1.3 & 40.5$\pm$0.7 &  3.8$\pm$0.5 & 100.0$\pm$0.8 & 140.3$\pm$ 1.3 & 434.5$\pm$ 3.9 & 346.4$\pm$ 2.9 & 32.5$\pm$ 0.4 & 23.6$\pm$ 0.4 &   6.2$\pm$0.4 &    55.9  &  14.78   \\
\enddata
\tablenotetext{a}{Observed flux ratios
  100$\cdot$I($\lambda$)/I(H$\beta$) are shown.} 
\tablenotetext{b}{The number of individual galaxy in the SHOC
  catalog. Letters `a' and `b' indicate separate regions within the
  same galaxy.}  
\tablenotetext{c}{Equivalent width for the H$\beta$ line in emission.}
\tablenotetext{d}{Flux for the H$\beta$ line is in units of
  10$^{-15}$\ erg\ s$^{-1}$cm$^{-2}$.} 
\tablecomments{The complete version of this table is in the electronic
edition of the Journal.  The printed edition contains only a sample.}
\end{deluxetable}

\begin{deluxetable}{lrrrrrrrrrrrrccc}
\rotate
\setlength{\tabcolsep}{0.02in}
\tabletypesize{\tiny \bf}
\tablecaption{Corrected Relative Emission Line Fluxes and Errors\tablenotemark{a}.
\label{t:Intens_corr}
}
\tablewidth{0pt}
\tablehead{
\colhead{SHOC\tablenotemark{b}} & \colhead{[O\,{\sc ii}]}& \colhead{H10}       &\colhead{H9}         &\colhead{H$\delta$} & \colhead{H$\gamma$} & \colhead{[O\,{\sc iii}]} & \colhead{H$\beta$}  & \colhead{[O\,{\sc iii}]} & \colhead{[O\,{\sc iii}]} & \colhead{H$\alpha$} & \colhead{[S\,{\sc ii}]} & \colhead{[S\,{\sc ii}]} & \colhead{[O\,{\sc iii}]} & \colhead{C(H$\beta$)\tablenotemark{c}} & \colhead{EW(abs)\tablenotemark{d}}  \\
\colhead{ID}                    & \colhead{3727 \AA}     &  \colhead{3798 \AA} &  \colhead{3835 \AA} &\colhead{4101 \AA}  &  \colhead{4340 \AA} &  \colhead{4363 \AA}      & \colhead{4861 \AA}  & \colhead{4959 \AA}       & \colhead{5007 \AA}       & \colhead{6563 \AA}  & \colhead{6716 \AA}      & \colhead{6731 \AA}      & \colhead{7320+7330 \AA}  & \colhead{}                             & \colhead{\AA}                      \\
\colhead{(1)}                   & \colhead{(2)}          &  \colhead{(3)}      &  \colhead{(4)}      & \colhead{(5)}      & \colhead{(6)}       & \colhead{(7)}            & \colhead{(8)}       & \colhead{(9)}            & \colhead{(10)}           & \colhead{(11)}      & \colhead{(12)}          & \colhead{(13)}          & \colhead{(14)}           & \colhead{(15)}                         & \colhead{(16)}
}
\startdata
1    &  263.4$\pm$ 5.5 &     \nodata   &    6.8$\pm$ 5.9 &   25.7$\pm$ 2.5 &   46.9$\pm$1.9 &    2.7$\pm$1.0 &  100.0$\pm$2.1 &  107.2$\pm$ 1.8 &  319.9$\pm$ 4.5 &  288.1$\pm$ 4.5 &   28.5$\pm$ 1.0 &   20.6$\pm$ 0.9 &   3.6$\pm$1.2 &   0.490$\pm$0.018 &  1.400$\pm$0.397   \\
2    &  175.7$\pm$ 3.1 &  7.6$\pm$ 3.5 &    8.2$\pm$ 3.5 &   26.5$\pm$ 2.0 &   46.9$\pm$1.0 &    7.0$\pm$0.5 &  100.0$\pm$1.0 &  174.3$\pm$ 1.2 &  533.4$\pm$ 3.5 &  282.9$\pm$ 2.3 &   16.2$\pm$ 0.8 &   12.5$\pm$ 0.8 &   4.6$\pm$1.0 &   0.060$\pm$0.009 &  3.300$\pm$0.439   \\
3    &  239.0$\pm$ 5.5 &  1.4$\pm$ 1.1 &    4.5$\pm$ 1.4 &   28.1$\pm$ 1.3 &   49.4$\pm$1.2 &    7.6$\pm$0.9 &  100.0$\pm$1.6 &  187.7$\pm$ 2.6 &  559.0$\pm$ 7.0 &  282.5$\pm$ 3.7 &   16.9$\pm$ 1.0 &   12.5$\pm$ 0.9 &   3.7$\pm$1.0 &   0.030$\pm$0.015 &  0.000$\pm$0.278   \\
4    &  249.5$\pm$ 9.5 &     \nodata   &       \nodata   &   24.7$\pm$ 8.2 &   44.6$\pm$3.9 &    3.5$\pm$1.5 &  100.0$\pm$4.7 &   89.6$\pm$ 2.8 &  273.5$\pm$ 7.3 &  282.9$\pm$ 8.3 &   32.0$\pm$ 1.4 &   22.2$\pm$ 1.2 &   8.7$\pm$2.8 &   0.000$\pm$0.030 &  2.600$\pm$0.667   \\
5    &  251.4$\pm$ 6.2 &     \nodata   &       \nodata   &   25.9$\pm$ 2.9 &   46.6$\pm$2.2 &    3.6$\pm$0.9 &  100.0$\pm$2.3 &  142.1$\pm$ 2.2 &  434.0$\pm$ 6.4 &  286.1$\pm$ 4.5 &   26.4$\pm$ 0.7 &   17.5$\pm$ 0.6 &   5.2$\pm$0.7 &   0.020$\pm$0.018 &  2.500$\pm$0.499   \\
6    &  259.6$\pm$ 6.2 &     \nodata   &    4.2$\pm$ 2.7 &   24.5$\pm$ 3.0 &   47.4$\pm$1.9 &    1.7$\pm$0.9 &  100.0$\pm$2.1 &  135.8$\pm$ 1.9 &  407.2$\pm$ 5.5 &  285.9$\pm$ 4.2 &   23.0$\pm$ 0.8 &   17.2$\pm$ 0.7 &   2.8$\pm$0.6 &   0.000$\pm$0.017 &  0.500$\pm$0.577   \\
7    &  286.8$\pm$ 7.8 &     \nodata   &    4.1$\pm$ 3.3 &   27.4$\pm$ 3.9 &   49.5$\pm$2.9 &    3.4$\pm$1.7 &  100.0$\pm$2.7 &   83.2$\pm$ 1.7 &  242.9$\pm$ 4.2 &  283.3$\pm$ 5.4 &   33.2$\pm$ 1.4 &   24.3$\pm$ 1.3 &   1.9$\pm$0.9 &   0.260$\pm$0.022 &  0.000$\pm$0.697   \\
\enddata
\tablenotetext{a}{Corrected intensity ratios
  100$\cdot$I($\lambda$)/I(H$\beta$) are shown.} 
\tablenotetext{b}{The number of individual galaxy in the SHOC
  catalog. Letters `a' and `b' indicate separate regions within the
  same galaxy.}  
\tablenotetext{c}{Extinction coefficient.}
\tablenotetext{d}{Equivalent width of underlying absorption in Balmer
  hydrogen lines.} 
\tablecomments{The complete version of this table is in the electronic
edition of the Journal.  The printed edition contains only a sample.}
\end{deluxetable}

\begin{deluxetable}{lrrrrrr}
\tabletypesize{\footnotesize \bf}
\tablecaption{Oxygen Abundances in SHOC galaxies
\label{t:Abun}
}
\tablewidth{0pt}
\tablehead{
\colhead{SHOC} & \colhead{$T_\mathrm{e}$(O{\sc iii})}& \colhead{$T_\mathrm{e}$(O{\sc ii})}&  \colhead{$N_\mathrm{e}$(S{\sc ii})}& \colhead{O$^{+}$/H$^{+}$}& \colhead{O$^{++}$/H$^{+}$}& \colhead{12+log(O/H)} \\
\colhead{ID}   & \colhead{K}                   &  \colhead{K}                 &  \colhead{cm$^{-3}$}          & \colhead{$\times$10$^5$} & \colhead{$\times$10$^5$}  &                       \\
\colhead{(1)}  & \colhead{(2)}                 &  \colhead{(3)}               &  \colhead{(5)}                & \colhead{(6)}            & \colhead{(7)}             & \colhead{(8)}
}
\startdata
1    &    10900$\pm$1300 &    11500$\pm$1400  &    37$\pm$   48 &    5.57$\pm$  2.02 &   8.47$\pm$  3.18 &     8.15$\pm$    0.12   \\
2    &    12800$\pm$ 400 &    12600$\pm$ 400  &   127$\pm$  110 &    2.68$\pm$  0.23 &   8.72$\pm$  0.73 &     8.06$\pm$    0.03   \\
3    &    13000$\pm$ 600 &    12700$\pm$ 700  &    64$\pm$   97 &    3.54$\pm$  0.46 &   8.90$\pm$  1.13 &     8.09$\pm$    0.04   \\
4    &    12700$\pm$2000 &    12600$\pm$2300  &     $<$10       &    3.81$\pm$  1.76 &   4.57$\pm$  2.09 &     7.92$\pm$    0.14   \\
5    &    10900$\pm$1000 &    11500$\pm$1100  &     $<$10       &    5.32$\pm$  1.42 &  11.49$\pm$  3.17 &     8.23$\pm$    0.09   \\
6    &     8800$\pm$1300 &    10100$\pm$1400  &    75$\pm$   62 &    9.25$\pm$  4.48 &  22.64$\pm$ 12.34 &     8.50$\pm$    0.18   \\
7    &    13100$\pm$2600 &    12800$\pm$3000  &    55$\pm$   70 &    4.15$\pm$  2.35 &   3.76$\pm$  2.09 &     7.90$\pm$    0.17   \\
8    &    11300$\pm$ 500 &    11800$\pm$ 600  &    47$\pm$   28 &    4.28$\pm$  0.63 &   9.47$\pm$  1.34 &     8.14$\pm$    0.05   \\
\enddata
\tablecomments{The complete version of this table is in the electronic
edition of the Journal.  The printed edition contains only a sample.}
\end{deluxetable}

\begin{deluxetable}{llcccl}
\tabletypesize{\bf}
\tablecaption{12+log(O/H) from SDSS data and from other sources
\label{tbl:others}
}
\tablewidth{0pt}
\tablehead{
\colhead{SHOC ID} & \colhead{Name} & \colhead{SDSS} & \colhead{Others} & \colhead{References}  & \colhead{Comment\tablenotemark{a}} \\
\colhead{(1)}     & \colhead{(2)}  & \colhead{(3)}  & \colhead{(4)}     & \colhead{(5)}  & \colhead{(6)}
}
\startdata
22   & HS 0029+1443     & 8.07$\pm$0.02  & 7.96$\pm$0.07 & 9   & N \\
220  & HS 0837+4717     & 7.62$\pm$0.01  & 7.63$\pm$0.03 & 6   & Y \\
254a & MRK 1416         & 7.92$\pm$0.02  & 7.86$\pm$0.02 & 5   & N \\
254b & MRK 1416         & 7.81$\pm$0.04  & 7.86$\pm$0.02 & 5   & N \\
261  & MRK 0116 (SE)    & 7.25$\pm$0.04  & 7.19$\pm$0.02 & 4   & N \\
261  & MRK 0116 (SE)    & 7.25$\pm$0.04  & 7.26$\pm$0.05 & 9   & N \\
280  & MRK 0022         & 8.02$\pm$0.02  & 8.00$\pm$0.01 & 5   & N \\
281a & MRK 1236         & 8.10$\pm$0.01  & 8.07$\pm$0.03 & 2   & N \\
\enddata
\tablenotetext{a}{
 Oxygen abundances calculated with (Y) or without (N) [O~{\sc ii}]
 $\lambda$3727 \AA; the set of numbers correspond to the
 truncation flag from column 7 in Table~\ref{tab:General}.} 
\tablenotetext{b}{Abundances were taken from \citet{Kniazev03a}.}
\tablerefs{(1) \citet{Denicolo};
	   (2) \citet{G00};
	   (3) \citet{G03};
	   (4) \citet{IT98};
	   (5) \citet{IT99};
	   (6) \citet{Kniazev00b};
	   (7) \citet{Mas94};
	   (8) \citet{Pust03};
	   (9) \citet{SK93}.
}
\end{deluxetable}

\setcounter{tab}{0}
\begin{deluxetable}{lllrccl}
\tabletypesize{\footnotesize \bf}
\tablecaption{General Parameters of Strong Narrow-Line
	      AGN and LINERs from SDSS DR1
\label{tab:AGN}
}
\tablewidth{0pt}
\tablehead{
\colhead{\#} & \colhead{SDSS name} & \colhead{Plate,MJD,Fiber} &
\colhead{$r$} & \colhead{$z$} & \colhead{Type} & \colhead{Comments} \\
	      & \colhead{(1)}      & \colhead{(2)}             &
\colhead{(3)} & \colhead{(4)} & \colhead{(5)}  & \colhead{(6)}
}
\startdata
\numb & SDSS J002119.69$+$003801.7 & 390 51900 454 & 18.84  & 0.23382 & LINER &                      \\
\numb & SDSS J002531.46$-$104022.0 & 653 52145 149 & 18.26  & 0.30349 &  AGN  &                      \\
\numb & SDSS J002837.83$-$095953.7 & 654 52146 353 & 16.48  & 0.04961 & LINER &                      \\
\numb & SDSS J002916.68$+$151101.6 & 417 51821 396 & 19.29  & 0.21595 & LINER &                      \\
\numb & SDSS J003356.44$-$004521.6 & 392 51793 251 & 17.47  & 0.17380 &  AGN  &                      \\
\numb & SDSS J004759.78$+$134433.7 & 419 51879   8 & 15.87  & 0.05657 &  AGN  &                      \\
\numb & SDSS J005051.46$-$084100.0 & 657 52177 327 & 17.37  & 0.09654 &  AGN  &                      \\
\numb & SDSS J005800.82$+$002845.2 & 395 51783 469 & 16.65  & 0.08018 & LINER &                      \\
\enddata
\end{deluxetable}

\setcounter{tab}{0}
\begin{deluxetable}{lllllll}
\tabletypesize{\footnotesize \bf}
\tablecaption{General Parameters of Additional
	      Strong Narrow-Line Galaxies with WR-features from SDSS DR1
\label{tab:WR}}
\tablewidth{0pt}
\tablehead{
\colhead{\#} & \colhead{SDSS name} & \colhead{Plate,MJD,Fiber} &
\colhead{$r$} & \colhead{$z$} & \colhead{WR\tablenotemark{a}} & \colhead{Comments} \\
\colhead{(1)} & \colhead{(2)} & \colhead{(3)} &
\colhead{(4)} & \colhead{(5)} & \colhead{(6)} & \colhead{(7)}
}
\startdata
\numb & SDSS J002837.83$-$095953.7 & 654 52146 353 & 16.48 &  0.04961 & ~3~ & LINER     \\
\numb & SDSS J003650.43$+$000201.9 & 392 51793 471 & 17.71 &  0.11169 & ~3~ &           \\
\numb & SDSS J003837.62$-$095217.9 & 655 52162 271 & 18.45 &  0.06532 & ~3~ &           \\
\numb & SDSS J014249.30$-$093626.8 & 664 52174 276 & 18.16 &  0.16094 & ~1~ &           \\
\numb & SDSS J023533.89$-$092147.3 & 455 51909 132 & 21.94 &  0.00458 & ~3~ &           \\
\numb & SDSS J024630.36$-$001431.5 & 409 51871 225 & 17.20 &  0.00970 & ~1~ &           \\
\numb & SDSS J032046.10$-$005100.0 & 413 51929  47 & 19.56 &  0.02172 & ~2~ &           \\
\numb & SDSS J085258.32$+$492733.8 & 551 51993 279 & 16.12 &  0.00957 & ~3~ &           \\
\enddata
\tablenotetext{a}{The Wolf-Rayet galaxy flag.  ``1'', ``2'', or
``3'' means that either only the ``blue'' bump, or only the ``red'' bump,
or both bumps are detected.  A cross identifies the WR galaxy 
from the list of suspected WR galaxies by SCP99.}
\end{deluxetable}

\end{document}